\documentclass[final,5p,times,twocolumn]{elsarticle}

\biboptions{sort&compress}

\usepackage{mathtools}
\usepackage{braket} 
\usepackage[dvipsnames]{xcolor}  
\usepackage{float}
\usepackage[T1]{fontenc}
\usepackage[utf8]{inputenc}
\usepackage{graphicx,color,rotating,pifont}
\usepackage{amsmath,amssymb}
\usepackage{mathrsfs}
\usepackage{mathtools}
\usepackage{subfigure}
\usepackage{siunitx}
\usepackage{bm}
\usepackage{ae}
\usepackage{dcolumn}
\usepackage{txfonts}
\usepackage{tensor}
\usepackage{braket}
\usepackage{booktabs}
\usepackage{xspace}
\usepackage{soul}
\usepackage{multirow}
\usepackage{tikz}
\usepackage{pgffor} 
\setstcolor{red}
\setul{}{.2ex} 
 
\usepackage[normalem]{ulem}

\usetikzlibrary{backgrounds}
\usetikzlibrary{shapes,matrix,trees}
\usetikzlibrary{arrows.meta}
\usetikzlibrary{positioning}			
\usetikzlibrary{calc,through}			
\usetikzlibrary{positioning,calc,shapes,decorations.pathreplacing,decorations.markings,decorations.pathmorphing}
\tikzset{
	vector/.style={decorate, decoration={snake,amplitude=2.5pt}, draw},
	provector/.style={decorate, decoration={snake,amplitude=2.5pt}, draw},
	antivector/.style={decorate, decoration={snake,amplitude=-2.5pt}, draw},
	fermion/.style={draw=black, postaction={decorate},
		decoration={markings,mark=at position .6 with {\arrow[draw=black]{>}}}},
           vL/.style={draw=ppurple, postaction={decorate},
		decoration={markings,mark=at position .6 with {\arrow[draw=ppurple]{>}}}},
	vLp/.style={draw=ppurple, postaction={decorate},
		decoration={markings,mark=at position .7 with {\arrow[draw=ppurple]{>}}}},
	NR/.style={draw=ggreen, postaction={decorate},
		decoration={markings,mark=at position .62 with {\arrow[draw=ggreen]{>}}}},	
	NRp/.style={draw=ggreen, postaction={decorate},
		decoration={markings,mark=at position .7 with {\arrow[draw=ggreen]{>}}}},	
	neutralino/.style={draw=black},
	fermionbar/.style={draw=black, postaction={decorate},
		decoration={markings,mark=at position .6 with {\arrow[draw=black]{<}}}},
	fermionnoarrow/.style={draw=black},
	gluon/.style={decorate, draw=black,
		decoration={coil,amplitude=4pt, segment length=5pt}},
	scalar/.style={dashed,draw=black, postaction={decorate},
		decoration={markings,mark=at position .55 with {\arrow[draw=black]{>}}}},
	scalarbar/.style={dashed,draw=black, postaction={decorate},
		decoration={markings,mark=at position .55 with {\arrow[draw=black]{<}}}},
	scalarnoarrow/.style={dashed,draw=black},
	electron/.style={draw=black, postaction={decorate},
		decoration={markings,mark=at position .55 with {\arrow[draw=black]{>}}}},
	bigvector/.style={decorate, decoration={snake,amplitude=4pt}, draw},
	photon/.style={decorate, draw=black,decoration={snake,amplitude=4pt, segment length=5pt} }
}

\usepackage[colorlinks,linkcolor=blue,anchorcolor=blue,citecolor=blue]{hyperref} 
\hypersetup{
   colorlinks=true, 
   linkcolor=blue,  
   citecolor=blue,  
   filecolor=blue,  
   urlcolor=blue    
} 

\DeclareMathSymbol{\NS}{\mathord}{AMSb}{"4E}

\DeclareSIUnit{\fm}{\femto\meter}

\newcommand{\beq}{\begin{equation}}
\newcommand{\eeq}{\end{equation}}
\newcommand{\beqn}{\begin{eqnarray}}
\newcommand{\eeqn}{\end{eqnarray}}
\newcommand{\bsub}{\begin{subequations}}
\newcommand{\esub}{\end{subequations}}
\newcommand{\bpm}{\begin{pmatrix}}
\newcommand{\epm}{\end{pmatrix}}

\newcommand\identity{1\kern-0.25em\text{l}}

\makeatletter

\newcommand{\Rmnum}[1]{\expandafter\@slowromancap\romannumeral #1@}
\makeatother

\begin{document}

\begin{frontmatter}
 
\title{Convergence of many-body truncations in the in-medium similarity renormalization group}
 
\author{L. H. Chen\fnref{1,2}}
\author{H. Hergert\fnref{3,4}}
 \author{J. M. Yao\fnref{1,2}\corref{cor1}}
    \ead{yaojm8@sysu.edu.cn}
    \cortext[cor1]{Corresponding author}

\address[1]{School of Physics and Astronomy, Sun Yat-sen University, Zhuhai 519082, P.R. China}  
  \address[2]{Guangdong Provincial Key Laboratory of Quantum Metrology and Sensing, Sun Yat-Sen University, Zhuhai 519082, China }  
\address[3]{Facility for Rare Isotope Beams, Michigan State University, East Lansing, Michigan 48824-1321, USA.}
\address[4]{Department of Physics \& Astronomy, Michigan State University, East Lansing, Michigan 48824-1321, USA.}

\date{\today}

\begin{abstract}
The in-medium similarity renormalization group (IMSRG) is a leading \textit{ab initio} method for describing atomic nuclei. Most applications truncate the flowing
operators at the normal-ordered two-body level, neglecting induced
higher-body terms. Convergence with increasing operator rank
remains poorly understood, particularly in strongly correlated
systems. We systematically assess the many-body truncation hierarchy in the single-reference and multi-reference IMSRG, retaining normal-ordered operators through the four-body and
three-body levels, respectively. Using an exactly solvable pairing-plus-particle-hole model, we quantify successive higher-body corrections over a broad range of interaction
strengths, spanning weakly to strongly correlated regimes. The multi-reference formulation exhibits faster and more robust
convergence than its single-reference counterpart, with the
advantage becoming most pronounced at strong coupling. In this regime, higher-rank irreducible density matrices play an essential
role in achieving convergence of the many-body expansion.
\end{abstract}


\end{frontmatter}

\section{Introduction}

 Solving the quantum many-body problem for a given Hamiltonian, particularly in the presence of strong correlations, is a central challenge in modern physics. In nuclear physics, the development of high-precision interactions within chiral effective field theory, which incorporates the symmetries of quantum chromodynamics, has increased the need for accurate \textit{ab initio} methods that quantitatively connect these interactions to nuclear observables. Substantial progress over the past decade has enabled such methods to solve the many-body Schrödinger equation in a systematically improvable manner across an increasingly broad range of nuclear masses~\cite{Hergert:2020abreview,Scalesi:2026lja}. Among these approaches, the in-medium similarity renormalization group (IMSRG)~\cite{Tsukiyama:2011PRL,Hergert:2016PR} has emerged as a particularly versatile framework, with applications spanning light to heavy nuclei~\cite{Stroberg:2021PRL,Hu:2022NP,Kuske:2025tsm}.

The IMSRG addresses the quantum many-body problem through a continuous sequence of unitary transformations that progressively decouple a chosen reference state from other states. 
While its time and memory requirements scale polynomially with the size of the single-particle basis and depend only indirectly on the particle number, IMSRG transformations generate operators of increasingly high rank, making it computationally demanding to retain them all. In most applications, the flowing operators are therefore truncated at the normal-ordered two-body (NO2B) level, defining the IMSRG(2) approximation~\cite{Hergert:2016PR}. Comparisons with other methods using the same Hamiltonian suggest that the omitted three- and higher-body operators have relatively small effects (about $1-2\%$) in the ground-state energies of near-closed-shell nuclei~\cite{Hergert:2020abreview}. Their effects may be more substantial in strongly correlated systems, particularly in open-shell deformed nuclei. Because retaining higher-rank operators is computationally demanding, systematic studies of convergence behaviors beyond the IMSRG(2) truncation remain scarce. Such studies are essential to quantify many-body truncation errors in IMSRG calculations.  

In recent years, the single-reference IMSRG (SR-IMSRG) has been extended to include all or selected contributions from normal-ordered three-body (NO3B) operators~\cite{Heinz:2021,Stroberg:2024,He:2024}. Calculations for closed-shell nuclei using chiral nuclear Hamiltonians in small model spaces have demonstrated subpercent errors in the ground-state energies. More recently, SR-IMSRG(4) was implemented for the Richardson pairing model~\cite{Hu:2026ilr}, revealing that contributions from operators beyond the NO3B level grow significantly with the pairing strength $g$ and reach a few percent of the total correlation energy when $g$ becomes comparable to the single-particle level spacing.

An alternative to employing a high-order SR-IMSRG($n$) truncation is to use a correlated reference state, leading to the multi-reference (MR) framework. Through generalized normal ordering with respect to such a reference state, the MR-IMSRG can incorporate pairing correlations~\cite{Hergert:2013PRL}, deformation~\cite{Yao:2020PRL}, and other static correlations~\cite{Gebrerufael:2017PRL} in open-shell nuclei more efficiently. This capability introduces additional terms involving higher-rank irreducible density matrices (also known as cumulants), but does not change the computational and memory scaling with the single-particle basis size $N$. Previous studies have shown that MR-IMSRG(2) can describe low-lying states of strongly deformed nuclei~\cite{Gebrerufael:2017PRL,Yao:2020PRL,Belley:2023lec,Zhou:2024vlt,Zhou:2026hya}, which are challenging for the single-reference formulation. Nevertheless, the size of the correlations omitted at the MR-IMSRG(2) level remains unclear. Quantifying corrections from explicit NO3B and higher-body terms in MR-IMSRG and comparing the convergence of the multi-reference and single-reference formulations as the many-body truncation order increases are therefore important steps toward establishing a controlled truncation hierarchy for (MR-)IMSRG calculations.

In this work, we present the first numerical implementation of the MR-IMSRG framework to explicitly retain all terms through the NO3B level, yielding the MR-IMSRG(3) approximation, and systematically benchmark the convergence of many-body truncations in both the SR-IMSRG and MR-IMSRG formulations. Using an exactly solvable pairing-plus-particle-hole (P3H) model over a broad range of interaction strengths~\cite{Davison:2023,Chen:2025ftv}, we quantify the corrections to the ground-state energy obtained by retaining operators of successively higher body rank. We find that the multi-reference formulation exhibits faster and more robust convergence than its single-reference counterpart, particularly in strongly correlated regimes.
 
\section{The MR-IMSRG(3) framework}
The P3H Hamiltonian consists of a one-body term with equally spaced single-particle levels and a two-body residual interaction of the form \cite{Davison:2023,Chen:2025ftv}
\beqn
\label{eq:H}
H_0
&=&
\delta \sum_i (|i|-1)a_i^\dagger a_i
-\frac{g}{2}\sum_{i,j>0}
a_i^\dagger a_{\bar i}^\dagger a_{\bar j}a_j
\nonumber\\
&-&
\frac{b}{2}\sum_{i,j\neq l>0}
\left(
a_i^\dagger a_{\bar i}^\dagger a_{\bar j}a_l
+
a_l^\dagger a_{\bar j}^\dagger a_{\bar i}a_i
\right).
\eeqn
Here, $\delta$ is the energy spacing between the single-particle levels labeled by $|i|=1,\ldots, L$, and $i$ and $\bar i$ denote states with opposite spin projections. The coupling constants $g$ and $b$ set the strengths of the pairing and pair-breaking interactions, respectively.

We construct a reference state with definite particle number by projecting a BCS wave function onto the $N$-particle subspace,
\begin{align}
\label{eq:PNPBCS-reference}
\ket{\Phi^{(N)}_{\rm ref}}
&=\frac{1}
{\sqrt{\mathcal N_N}}  \hat{P}^N
\prod_{p=1}^{L}
\left(u_p+\varv_p a_p^\dagger a_{\bar p}^\dagger\right)
\ket{0},
\end{align}
where $\hat{P}^{N}$ is the particle-number projection
operator~\cite{Ring:1980}, and $\mathcal{N}_N$ ensures that the
reference state is normalized. The occupation amplitudes $\varv_p$ for each level $p$ are determined by minimizing the projected energy within the variation-after-projection (VAP) scheme, with $u_p=\sqrt{1-\varv_p^2}$. The reference state (\ref{eq:PNPBCS-reference}) is referred to as PNP-BCS state hereafter. Further details are provided in the Appendix.

Using generalized normal ordering and Wick’s theorem~\cite{Kutzelnigg1997,Kong2010,Hergert:2017PS}, we express the Hamiltonian in Eq.~(\ref{eq:H}) in normal-ordered form with respect to the correlated reference state $\ket{\Phi^{(N)}_{\rm ref}}$. Its general normal-ordered expansion reads
\begin{align}
H
={}&
E
+\sum_{ij}f^i_j\left\{A^i_j\right\}
+\frac{1}{(2!)^2}\sum_{ijkl}
\Gamma^{ij}_{kl}\left\{A^{ij}_{kl}\right\}
\nonumber\\
&+\frac{1}{(3!)^2}\sum_{ijklmn}
W^{ijk}_{lmn}\left\{A^{ijk}_{lmn}\right\}
\nonumber\\
&+\frac{1}{(4!)^2}\sum_{ijklpqrs}
X^{ijkl}_{pqrs}\left\{A^{ijkl}_{pqrs}\right\}
+\cdots,
\label{eq:normal-ordered-Hamiltonian}
\end{align}
where $E$, $f$, $\Gamma$, $W$, and $X$ are the coefficients of the zero-, one-, two-, three-, and four-body terms, respectively. The MR-IMSRG(3) truncation retains normal-ordered operators through the three-body level, but we also implement the SR-IMSRG(4) truncation, which retains operators through the NO4B level using a Slater determinant reference state (see Appendix). Here, we have defined
$A^{i\cdots j}_{m\cdots n}
\equiv a^\dagger_i\cdots a^\dagger_j a_n\cdots a_m$,
and we use curly braces $\{\cdots\}$ to denote generalized normal ordering with respect to the reference state $\ket{\Phi^{(N)}_{\rm ref}}$. By construction, every normal-ordered operator of nonzero body rank has a vanishing expectation value in the reference state:
\begin{equation}
\bra{\Phi^{(N)}_{\rm ref}}
\left\{A^{i_1\cdots i_k}_{j_1\cdots j_k}\right\}
\ket{\Phi^{(N)}_{\rm ref}}
=0,
\qquad k\geq 1.
\end{equation}

We evolve the Hamiltonian in Eq.~(\ref{eq:normal-ordered-Hamiltonian}) through a continuous unitary transformation governed by the flow equation
\begin{equation}
\label{eq:SRG_flow_equation}
\frac{dH(s)}{ds}
=
[\eta(s),H(s)],
\end{equation}
where $s$ is the flow parameter and $\eta(s)$ is an anti-Hermitian generator. The initial condition is $H(0)=H_0$. Because $H_0$ contains at most two-body interactions, the higher-body terms in (\ref{eq:normal-ordered-Hamiltonian}) vanish initially: $W(0)=X(0)=\cdots=0$.

In the flow equation (\ref{eq:SRG_flow_equation}), we employ the Brillouin generator~\cite{Hergert:2017PS} at the NO3B level, which is given by
\begin{align}
\eta^{\mathrm{B}}(s)
={}&
\sum_{ij}\eta^i_j(s)\left\{A^i_j\right\}
+\frac{1}{4}\sum_{ijkl}
\eta^{ij}_{kl}(s)\left\{A^{ij}_{kl}\right\}
\nonumber\\
&+\frac{1}{36}\sum_{ijklmn}
\eta^{ijk}_{lmn}(s)\left\{A^{ijk}_{lmn}\right\},
\label{eq:Brillouin-generator}
\end{align}
with the matrix elements defined by
\begin{equation}
\eta^{i_1\cdots i_k}_{j_1\cdots j_k}(s)
\equiv
\bra{\Phi^{(N)}_{\mathrm{ref}}}
\left[
H(s),
\left\{A^{i_1\cdots i_k}_{j_1\cdots j_k}\right\}
\right]
\ket{\Phi^{(N)}_{\mathrm{ref}}}.
\label{eq:Brillouin-matrix-elements}
\end{equation}
The Brillouin generator is thus determined directly by the energy gradients associated with infinitesimal unitary transformations, without requiring energy denominators. Detailed expressions for its matrix elements, defined in Eq.~(\ref{eq:Brillouin-generator}), are provided in the Supplemental Material (SM).

Evaluating the commutator in Eq.~(\ref{eq:SRG_flow_equation})
yields a coupled set of ordinary differential equations for the
normal-ordered components $\{E,f,\Gamma,W,\ldots\}$. We derive
these equations automatically using our newly developed Python
package \texttt{qcombo}~\cite{Chen:2026qcombo}. Explicit expressions
for the MR-IMSRG(3) flow equations are provided in that reference.

The flow equations involve irreducible density matrices
$\lambda^{(k)}$ of the reference state. They are defined recursively in terms of the $k-$body density matrices $\rho^{(k)}$ as
\begin{equation}
\label{eq:irreducible_density}
\lambda^{(k)}
=
\rho^{(k)}
-
\sum_{\substack{
n_1+\cdots+n_m=k\\
m\geq 2,\;1\leq n_1\leq\cdots\leq n_m
}}
\mathcal{A}
\left[
\lambda^{(n_1)}
\cdots
\lambda^{(n_m)}
\right],
\end{equation}
where the operator $\mathcal{A}$ generates the distinct antisymmetrized terms
without double counting. The $k$-body density matrix is defined as
\begin{equation}
\label{eq:RDM-definition}
\rho^{i_1\cdots i_k}_{j_1\cdots j_k}
\equiv
\bra{\Phi^{(N)}_{\mathrm{ref}}}
A^{i_1\cdots i_k}_{j_1\cdots j_k}
\ket{\Phi^{(N)}_{\mathrm{ref}}}.
\end{equation}
For example, the $k=1$ and $2$ irreducible density matrices have the form
\bsub
\begin{align}
\lambda^i_j
&=
\rho^i_j,
\\
\lambda^{ij}_{kl}
&=
\rho^{ij}_{kl}
-
\lambda^i_k\lambda^j_l
+
\lambda^i_l\lambda^j_k.
\end{align}
\esub
The MR-IMSRG(3) flow equations require irreducible densities through rank $k=5$ \cite{Kutzelnigg1997,Kong2010,Hergert:2017PS}.
Irreducible densities of rank $k\geq2$ encode correlations in the reference state. For a single Slater determinant, as used in the SR-IMSRG, all $\lambda^{(k)}$ with $k\geq2$ vanish. 
For an $N$-particle system, the many-body densities $\rho^{(k)}$ vanish for $k>N$. The corresponding irreducible densities need not vanish, however, as they cancel antisymmetrized products of lower-rank irreducible densities in the cumulant expansion.
While a variety of approximations for the $\lambda^{(k)}$ have been described in the literature, e.g.,  so-called reconstructions in terms of $\lambda^{(1)}$ and $\lambda^{(2)}$ \cite{Valdemoro:1992yg,Colmenero:1993rm,Colmenero:1993ty,Nakatsuji:1996qf,Mazziotti:2007ly,kong2011_cumulant,Lackner:2015ul}, 
we work with the original definition \eqref{eq:irreducible_density}
in the MR-IMSRG(3) formalism and include all contributions involving irreducible densities through $\lambda^{(5)}$.

\section{Results and discussion} 
In this section, we systematically assess the accuracy of truncations in the MR-IMSRG(3) flow equations and generators by varying the maximum body rank of the retained operators and the maximum rank of the irreducible densities $\lambda^{(k)}$. As an illustrative example, we consider $N=4$ particles in $L=4$ doubly degenerate single-particle levels of the P3H model, representing an open-shell system. We measure energies in units of the level spacing $\delta$, so that the dimensionless interaction strengths in the Hamiltonian (\ref{eq:H}) are $g/\delta$ and $b/\delta$; equivalently, we set $\delta=1$. We benchmark the MR-IMSRG ground-state energies against exact results obtained by full configuration-interaction (FCI) diagonalization of the Hamiltonian.  

\begin{figure}
    \centering
    \includegraphics[width=\linewidth]{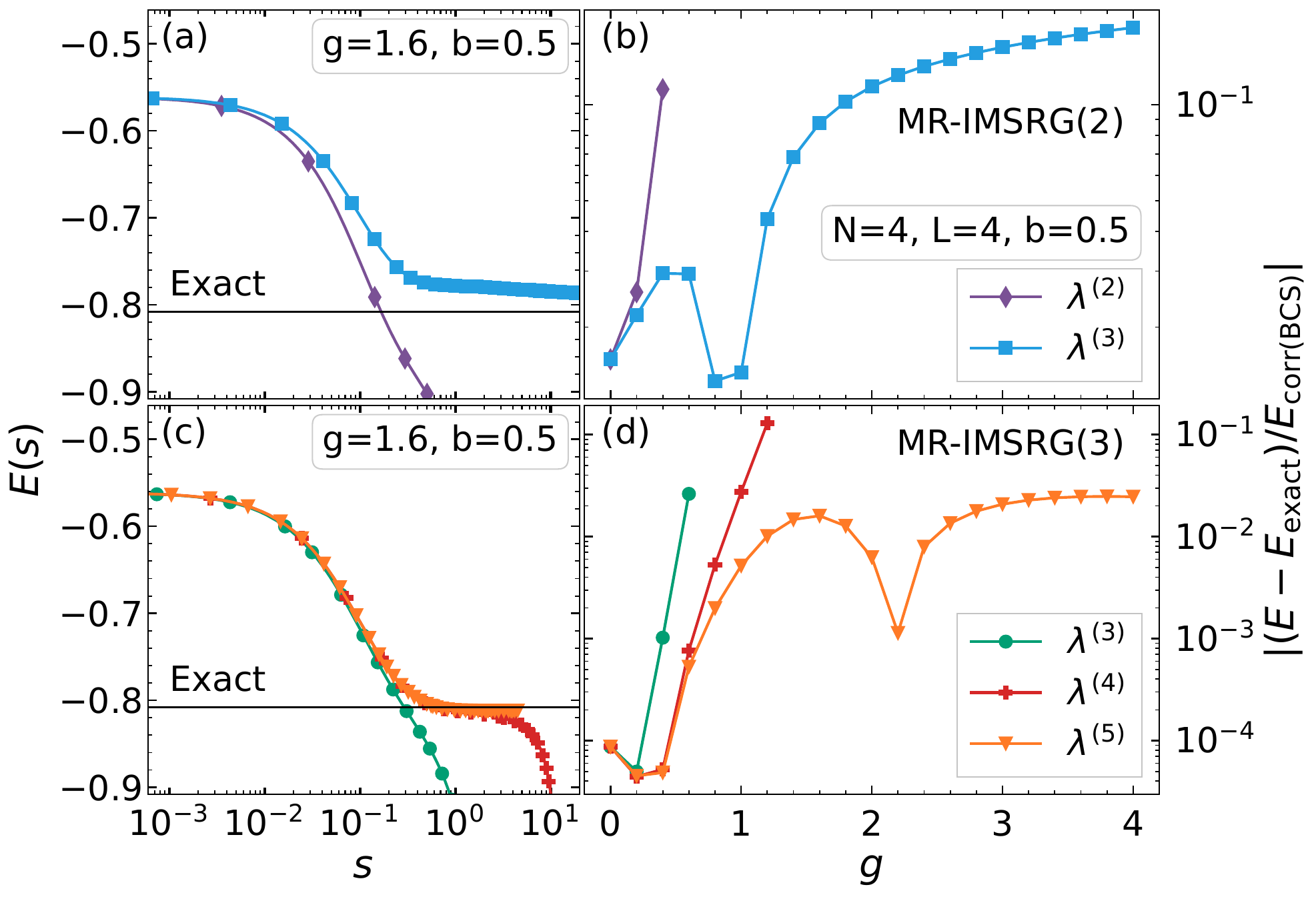}
    \caption{Ground-state energy as a function of the flow parameter $s$ [(a) and (c)] and relative error $\eta$, defined in Eq.~(\ref{eq:relative_error}), as a function of the pairing strength $g$ [(b) and (d)] for the P3H model with $N=4$, $L=4$, and $b=0.5$. The upper and lower panels show MR-IMSRG(2) and MR-IMSRG(3) results, respectively, with irreducible density matrices retained through different ranks. The horizontal black lines indicate the exact ground-state energies. Missing data points denote calculations that failed to converge. The dip in the relative error reflects a sign change in the energy error.}
    \label{fig:convergence_MRIMSRG_lambda}
\end{figure}

For illustration, we take $b=0.5$ and vary the pairing strength $g$ to probe different degrees of many-body correlations. The correlated reference is taken as the PNP-BCS state (\ref{eq:PNPBCS-reference}). Figure~\ref{fig:convergence_MRIMSRG_lambda}(a) and (c) shows the evolution of the ground-state energy with the flow parameter in MR-IMSRG(2) and MR-IMSRG(3), with the irreducible density matrices retained through different ranks. In MR-IMSRG(2), irreducible densities up to $\lambda^{(3)}$ enter the equations. 
As shown in Fig.~\ref{fig:convergence_MRIMSRG_lambda}(a), including the $\lambda^{(3)}$ terms in the generator $\eta$ is essential for convergence. In contrast, the $\lambda^{(3)}$ contribution to the flow equation for the zero-body term has a negligible effect on convergence. In MR-IMSRG(3), contributions from irreducible densities up to $\lambda^{(5)}$ must be retained to achieve convergence for $g>1.2$.

To assess the accuracy of truncations in the IMSRG($n$) flow equations, we define the normalized energy error $\eta$ as the deviation from the exact result normalized to the correlation energy beyond the reference state:
\begin{equation}
\label{eq:relative_error}
\eta\equiv
\frac{E-E_{\mathrm{exact}}}
{E_{\mathrm{ref}}-E_{\mathrm{exact}}},
\end{equation}
where $E$ is the IMSRG($n$) ground-state energy, $E_{\mathrm{ref}}$ is the reference-state energy, and $E_{\mathrm{exact}}$ is the exact ground-state energy obtained by full configuration interaction (FCI) diagonalization. The denominator represents the correlation energy that remains to be recovered beyond the reference state.

The results are displayed in  Fig.~\ref{fig:convergence_MRIMSRG_lambda}(b) and (d). We see that higher-rank irreducible densities become increasingly important for the accuracy of the final energy as
the pairing strength grows. This trend is consistent with the
pronounced increase in their traces and Frobenius norms
(see the Appendix). Although terms involving $\lambda^{(4)}$ and
$\lambda^{(5)}$ make relatively small contributions in a
perturbative treatment of the MR-IMSRG(3), their effects can
be amplified during the flow, altering its convergence behavior.
When all irreducible-density contributions required at each
operator truncation level are retained, the normalized energy
error in the MR-IMSRG(2) increases from a few percent to about
$10\%$ as $g$ increases from 0 to 4. Over the same range, the
MR-IMSRG(3) error increases from approximately $0.01\%$ to
about $1\%$.

\begin{figure}[bt]
\centering
\includegraphics[width=0.85\columnwidth]{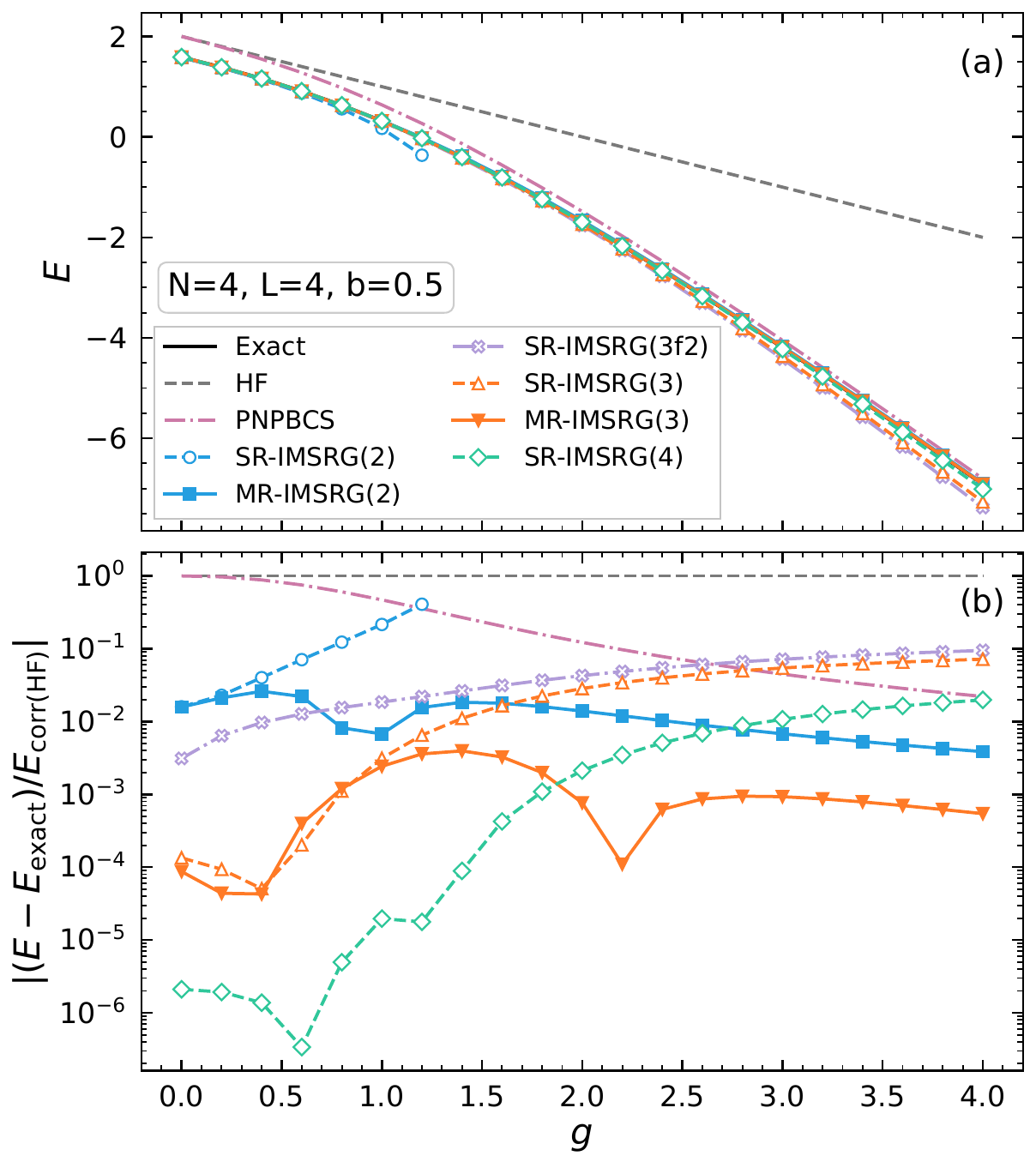}
\caption{(a) Ground-state energy as a function of the pairing strength $g$ from different calculations, including Hartree-Fock (HF), PNP-BCS (VAP), SR-IMSRG(n), MR-IMSRG(n), and FCI(Exact). (b) Corresponding absolute deviations from the exact energy, normalized by the HF correlation energy. 
}
\label{fig:MRvsSR}
\end{figure}

Furthermore, we compare the ground-state energies of the P3H model obtained using Hartree-Fock (HF), PNP-BCS (VAP), SR-IMSRG($n$), MR-IMSRG($n$), and FCI over a range of pairing strengths $g$. We also include results from the SR-IMSRG(3f2) scheme developed in Ref.~\cite{He:2024}, which approximately accounts for the effects of intermediate three-body operators generated in nested commutators. All MR-IMSRG calculations retain the full set of irreducible-density contributions required at their respective truncation levels. The energies are shown in Fig.~\ref{fig:MRvsSR}(a). Since the IMSRG results are close to the exact energies, their deviations are more clearly displayed in panel (b), which shows the normalized absolute errors
$|\eta|$ in (\ref{eq:relative_error}), where $E_{\rm ref}$ is defined as the HF energy. With this normalization, the HF curve lies at unity.

At weak coupling, SR-IMSRG($n$) and MR-IMSRG($n$) exhibit comparable accuracy. As $g$ increases, however, the SR-IMSRG truncation errors grow significantly, consistent with the findings of Ref.~\cite{Hu:2026ilr}. In particular, the SR-IMSRG(2) error increases dramatically, and the flow even diverges for the strong coupling region with $g>1.2$. This behavior reflects the increasing importance of induced three- and higher-body operators at stronger coupling. Retaining operators of higher body rank systematically improves the results: SR-IMSRG(3) reduces the error and extends the range of coupling strengths over which the flow converges, while SR-IMSRG(4) further improves the accuracy. The SR-IMSRG(3f2) error generally lies between those of SR-IMSRG(2) and SR-IMSRG(3), remaining closer to the former at weak coupling and approaching the latter as $g$ increases.

In contrast, the normalized errors of the MR-IMSRG calculations remain nearly constant or decrease as $g$ increases, becoming several orders of magnitude smaller than those of their SR-IMSRG counterparts. This behavior demonstrates the effectiveness of MR-IMSRG in the strongly correlated regime of the P3H model. Its advantage stems from the pairing correlations already incorporated in the PNP-BCS reference, which must instead be generated dynamically during the SR-IMSRG flow. At a fixed pair-breaking strength $b$, pairing becomes increasingly dominant as $g$ grows, and the PNP-BCS reference approaches the exact ground state. 
Nevertheless, MR-IMSRG(3) is systematically more accurate than MR-IMSRG(2), underscoring the importance of explicitly retaining NO3B operators even with a correlated reference. In the strong-pairing regime, MR-IMSRG(2) outperforms SR-IMSRG(3) for $g\gtrsim1.8$, with normalized errors generally below $1\%$, whereas MR-IMSRG(3) outperforms SR-IMSRG(4) for $g\gtrsim2.0$, with normalized errors generally below $0.1\%$.
 
\begin{figure}[bt]
    \centering
    \includegraphics[width=\linewidth]{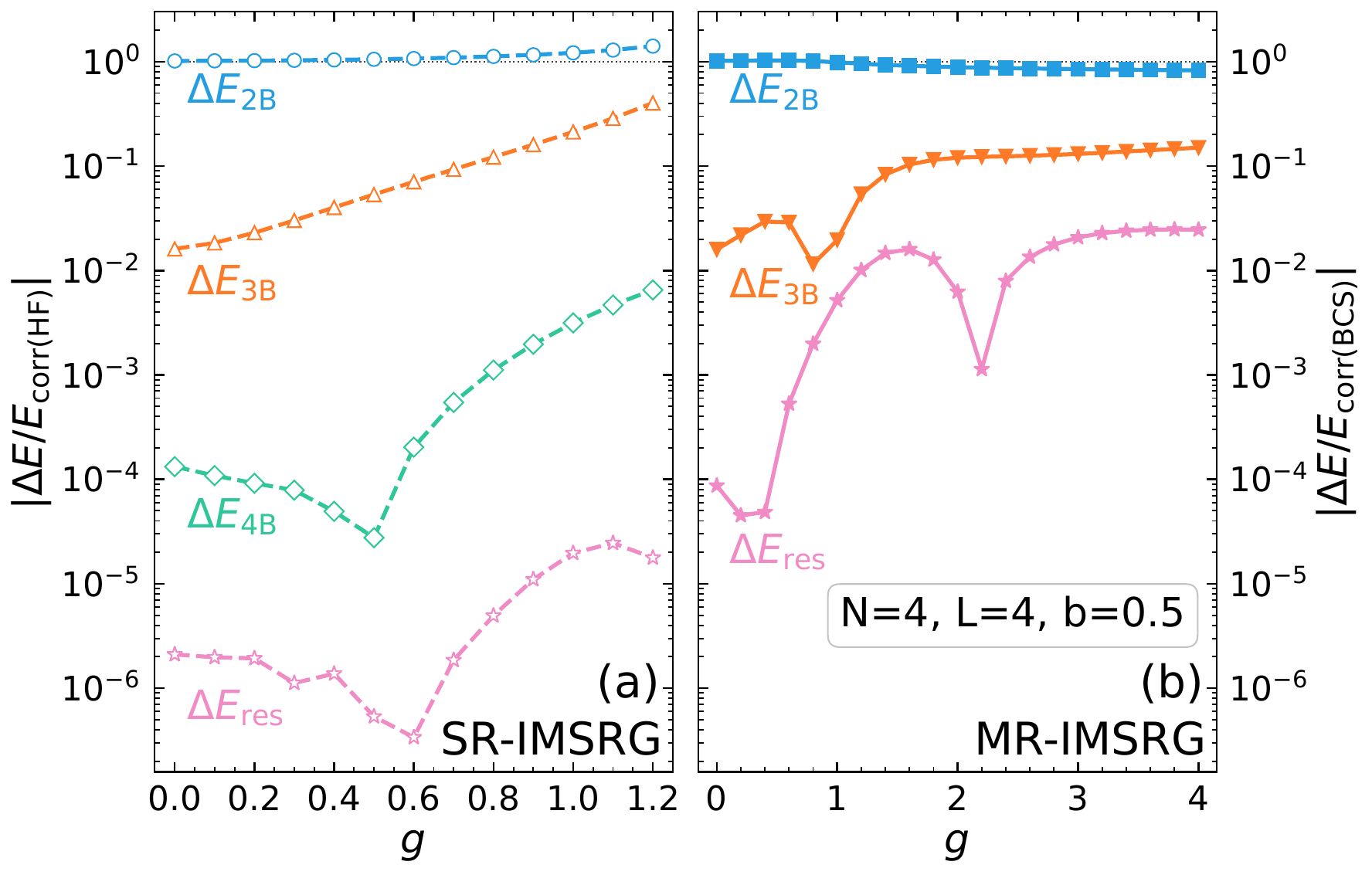}
    \caption{
    Corrections $\Delta E$ to the P3H ground-state energy from successive IMSRG truncation levels for $N=4, L=4$, and $b=0.5$. (a) SR-IMSRG corrections obtained by extending the operator truncation from NO2B through NO4B, normalized by the HF correlation energy. (b) Corresponding MR-IMSRG corrections obtained by extending the truncation from NO2B to NO3B, normalized by the PNP-BCS correlation energy. 
    }
    \label{fig:SR_MR_truncation}
\end{figure}

To quantify the hierarchy of energy corrections as the IMSRG
operator truncation is extended beyond NO2B, we decompose the
exact ground-state energy as
\begin{equation}
E_{\mathrm{exact}}
=
E_{\mathrm{ref}}
+\sum_{n=2}^{n_{\max}}\Delta E_{n\mathrm{B}}
+E_{\mathrm{res}},
\label{eq:energy-decomposition}
\end{equation}
where the incremental corrections and residual are defined by
\begin{subequations}
\label{eq:incremental-corrections}
\begin{align}
\Delta E_{2\mathrm{B}}
&=
E_{\mathrm{IMSRG}(2)}-E_{\mathrm{ref}},\\
\Delta E_{n\mathrm{B}}
&=
E_{\mathrm{IMSRG}(n)}
-E_{\mathrm{IMSRG}(n-1)}, \\
E_{\mathrm{res}}
&=
E_{\mathrm{exact}}
-E_{\mathrm{IMSRG}(n_{\max})}.
\end{align}
\end{subequations}
Thus, $\Delta E_{2\mathrm{B}}$ is the correlation energy recovered
at the NO2B level, while $\Delta E_{n\mathrm{B}}$ for $n\geq3$  measure changes between complete IMSRG($n$) and IMSRG($n-1$) calculations; they are not additive contributions from individual operator ranks evaluated along a common flow. The residual $E_{\mathrm{res}}$
measures the remaining discrepancy with the exact energy.
We use $n_{\max}=4$ for SR-IMSRG and $n_{\max}=3$ for MR-IMSRG,
with $E_{\mathrm{ref}}$ corresponding to the HF and PNP-BCS
reference energies, respectively.

Figures~\ref{fig:SR_MR_truncation}(a) and (b) show the energy contributions defined in Eq.~(\ref{eq:energy-decomposition}) for SR-IMSRG($n$) and MR-IMSRG($n$), respectively. The SR-IMSRG results are restricted to $g\leq1.2$, since SR-IMSRG(2) fails to converge beyond this value. Both formulations exhibit a hierarchy of successively smaller corrections as the retained operator body rank increases, supporting the use of finite-rank truncations within the convergence ranges examined.

In the SR formulation, successive corrections differ by approximately two orders of magnitude at weak coupling, but this hierarchy becomes less pronounced as $g$ increases.
In particular, $|\Delta E_{3\mathrm{B}}|/E_{\mathrm{corr(HF)}}$ grows from approximately $1\%$ to $40\%$, while the normalized four-body correction increases from about $0.01\%$ to nearly $1\%$, consistent with the finding in Ref. \cite{Hu:2026ilr}.
These growing contributions reflect the increasing importance of induced many-body operators as pairing correlations develop beyond the HF reference. Omitting these operators leads to larger truncation errors and contributes to the loss of convergence of SR-IMSRG(2) at stronger coupling.

The MR formulation exhibits a similar hierarchy but a qualitatively different dependence on $g$. The MR-IMSRG(2) correction recovers most of the correlation energy missing from the PNP-BCS reference. The additional three-body correction is typically about an order of magnitude smaller, and the residual discrepancy with the exact energy is even smaller. Following an initial variation, the normalized corrections tend to saturate as $g$ increases, while the hierarchy remains well separated even in the strong-pairing regime. 
This behavior reflects the incorporation of the dominant $g$-dependent pairing correlations directly into the PNP-BCS reference, leaving corrections with a weaker dependence on $g$ as well as correlations from the pair-breaking term to be generated during the MR-IMSRG flow. These results highlight the efficiency of the MR framework in capturing strong pairing correlations.  Since the two panels normalize the corrections to the correlation energies beyond their respective reference states, their vertical scales cannot be used to directly compare the absolute energy corrections.

Finally, we verify that the above findings also hold for $b=0$, the corresponding results are provided in the SM. In this case, PNP-BCS already yields a near-exact solution, so the additional effects of the MR-IMSRG flow are much smaller than for $b\neq0$. All other qualitative conclusions remain unchanged.

\section{Summary}
We have implemented the MR-IMSRG(3) framework for the pairing-plus-particle-hole model and systematically investigated the convergence of many-body truncations in the single-reference (SR) and multi-reference (MR) IMSRG formulations, retaining normal-ordered operators through the four-body (NO4B) and three-body (NO3B) levels, respectively. Benchmarks against exact solutions over a range of pairing strengths quantify the energy corrections obtained by successively increasing the retained operator body rank. The results show that the MR formulation generally exhibits faster and more robust convergence than its SR counterpart, particularly in the strong-pairing regime. These findings offer important guidance for assessing many-body truncation errors in IMSRG calculations of open-shell nuclei with modern chiral interactions. They also illustrate the merit of incorporating collective correlations, e.g., from the spontaneous breaking of gauge or rotational symmetries, into the reference state while letting the MR-IMSRG evolution capture dynamical few-particle correlations. Extensions to realistic nuclear systems are in progress.

\section*{Acknowledgments} 
 L.H. C. and J.M.Y. acknowledge the support by the National Natural Science Foundation of China (Grant No.  12375119).  H.H. acknowledges the support of Michigan State University.
  
\section*{Data availability}

The data supporting the findings of this study are openly
available on Zenodo at~\cite{IMSRG_zenodo_2026}.

\section*{Appendix: Construction of reference states}
\label{app:PNPBCS}

For SR-IMSRG($n$), we use a Slater determinant as the reference state for the $N$-particle system. For simplicity, we refer to it as the HF state, although it is fixed to the lowest-energy configuration at $g=b=0$ rather than optimized for the full Hamiltonian at each pair of $g$ and $b$ values. The HF state reads
\begin{equation}
\ket{\Phi_{\mathrm{HF}}}
=
\prod_{i=1}^{N}a_i^\dagger\ket{0},
\end{equation}
where the product runs over the occupied levels. For the four-particle P3H system considered here, this state reads
\begin{equation}
\ket{\Phi_{\mathrm{HF}}}
=
a_1^\dagger a_{\bar{1}}^\dagger
a_2^\dagger a_{\bar{2}}^\dagger\ket{0}.
\label{eq:HF-reference-P3H}
\end{equation}
In the above basis, the one-body density matrix is diagonal:
\begin{equation}
\rho^i_j\equiv\bra{\Phi_{\mathrm{HF}}} a^\dagger_i a_j\ket{\Phi_{\mathrm{HF}}}=n_i\delta^i_j,
\label{eq:HF-one-body-density}
\end{equation}
where $n_i=1$ for the occupied (hole) states
$(1,\bar{1},2,\bar{2})$ and $n_i=0$ for the unoccupied (particle) states. The one-body irreducible density equals the one-body density matrix, $\lambda^i_j=\rho^i_j$, while all irreducible densities of rank $k\geq2$ vanish.

For MR-IMSRG($n$), we use a particle-number-projected BCS (PNP-BCS) reference state,
\begin{align}
\ket{\Phi^{(N)}_{\mathrm{ref}}}
&=
\frac{1}{\sqrt{\mathcal N_N}}\hat P^N
\prod_{p=1}^{L}
\left(u_p+\varv_p\hat P_p^\dagger\right)\ket{0},
\qquad
\hat P_p^\dagger=a_p^\dagger a_{\bar p}^\dagger,
\end{align}
where $\hat P^N$ projects onto the $N$-particle subspace, $\mathcal N_N$ is the normalization factor, and $\hat P_p^\dagger$ creates a pair in the doubly degenerate level $p$. We parametrize the real amplitudes as $u_p=\cos\theta_p$ and $\varv_p=\sin\theta_p$, which automatically ensures $u_p^2+\varv_p^2=1$. 

For an even particle number $N=2M$, the projected reference state can be written as
\begin{equation}
\ket{\Phi^{(N)}_{\mathrm{ref}}(\theta)}
=
\frac{1}{\sqrt{\mathcal N_N(\theta)}}
\sum_{1\leq k_1<\cdots<k_M\leq L}
C_{k_1\cdots k_M}(\theta)
\hat P_{k_1}^\dagger\cdots\hat P_{k_M}^\dagger\ket{0},
\label{eq:app-projected-state}
\end{equation}
where
\begin{equation}
C_{k_1\cdots k_M}(\theta)
=
\prod_{\alpha=1}^{M}\varv_{k_\alpha}
\prod_{\ell\notin\{k_1,\ldots,k_M\}}u_\ell,
\label{eq:app-projected-coefficients}
\end{equation}
and
\begin{equation}
\mathcal N_N(\theta)
=
\sum_{1\leq k_1<\cdots<k_M\leq L}
C_{k_1\cdots k_M}^{\,2}(\theta).
\end{equation}
The variational parameters $\theta=\{\theta_p\}$ are determined by minimizing the projected energy,
\begin{equation}
E^{(N)}_{\rm ref}(\theta)
=
\bra{\Phi^{(N)}_{\mathrm{ref}}(\theta)}
H_0
\ket{\Phi^{(N)}_{\mathrm{ref}}(\theta)}.
\label{eq:app-PNPBCS-energy}
\end{equation}

For $N=4$ and $L=4$, each configuration contains two occupied pair orbitals, giving
\begin{equation}
\ket{\Phi^{(4)}_{\mathrm{ref}}}
=
\frac{1}{\sqrt{\mathcal N}}
\sum_{1\leq p<q\leq4}
C_{pq}\hat P_p^\dagger\hat P_q^\dagger\ket{0},
\label{eq:app-PNPBCS-four-particles}
\end{equation}
with
\begin{equation}
C_{pq}
=
\varv_p\varv_q\prod_{r\ne p,q}u_r,
\qquad
\mathcal N
=
\sum_{1\leq p<q\leq4}C_{pq}^{\,2}.
\label{eq:app-Cpq}
\end{equation}
Hereafter, $u_p$ and $\varv_p$ denote the optimized amplitudes, and we define $C_{qp}=C_{pq}$ and $C_{pp}=0$.

The $k$-body density matrix of this reference is
\begin{equation}
\rho^{i_1\cdots i_k}_{j_1\cdots j_k}
\equiv
\bra{\Phi^{(4)}_{\mathrm{ref}}}
a_{i_1}^\dagger\cdots a_{i_k}^\dagger
a_{j_k}\cdots a_{j_1}
\ket{\Phi^{(4)}_{\mathrm{ref}}}.
\label{eq:app-RDM-definition}
\end{equation}
All elements through rank four can be expressed directly in terms of $C_{pq}$. Below, Latin indices $p,q,r,s$ label pair orbitals, Greek indices label individual single-particle states, and $|\alpha|$ denotes the pair orbital containing state $\alpha$.

The one-body density matrix is diagonal:
\begin{equation}
\rho^\alpha_\beta
=
\delta_{\alpha\beta}n_{|\alpha|},
\qquad
n_p
=
\frac{1}{\mathcal N}\sum_{r=1}^{4}C_{pr}^{\,2}.
\label{eq:app-rho1}
\end{equation}
Both members of a time-reversed pair have the same occupation,
$\rho^p_p=\rho^{\bar p}_{\bar p}=n_p$.
The single-particle states forming these pairs therefore constitute a natural-orbital basis, up to reordering and rotations within degenerate subspaces.

For arbitrary single-particle indices, the two-body density matrix reads
\begin{align}
\rho^{\alpha\beta}_{\gamma\delta}
={}&
\delta_{\alpha\bar{\beta}}\delta_{\gamma\bar{\delta}}
\sigma(\alpha)\sigma(\gamma)
\frac{1}{\mathcal N}
\sum_{r=1}^{4}C_{|\alpha|r}C_{|\gamma|r}
\nonumber\\
&+
\left(
\delta_{\alpha\gamma}\delta_{\beta\delta}
-
\delta_{\alpha\delta}\delta_{\beta\gamma}
\right)
\frac{C_{|\alpha|\,|\beta|}^{\,2}}{\mathcal N},
\label{eq:app-rho2-pair}
\end{align}
where $\sigma(p)=+1$ and $\sigma(\bar p)=-1$.
The first term describes complete-pair transfer, with the sign factors ensuring antisymmetry within each index set. The second describes joint occupation of states in different pair orbitals and vanishes when $|\alpha|=|\beta|$ because $C_{pp}=0$.

A nonzero three-body density-matrix element requires each index set to contain one complete pair and one state from another pair orbital. In the ordering used below,
\begin{equation}
\rho^{p\bar p\,\alpha}_{q\bar q\,\beta}
=
\delta_{\alpha\beta}
\frac{C_{p|\alpha|}C_{q|\alpha|}}{\mathcal N}.
\label{eq:app-rho3}
\end{equation}
Annihilating three particles leaves the time-reversed partner of the unpaired index, so the overlap of the remaining states requires $\alpha=\beta$.

For the four-body density matrix, each index set must contain two complete pairs. Annihilating all four particles gives
\begin{equation}
\rho^{p\bar p\,q\bar q}_{r\bar r\,s\bar s}
=
\frac{C_{pq}C_{rs}}{\mathcal N}.
\label{eq:app-rho4}
\end{equation}
The upper and lower pair configurations may be identical, partially overlapping, or disjoint.

Equations~\eqref{eq:app-rho3}--\eqref{eq:app-rho4} determine all nonzero density-matrix elements for the $N=4$ case. Other index orderings follow from antisymmetry within the upper and lower index sets:
\begin{equation}
\rho^{i_{\pi(1)}\cdots i_{\pi(k)}}_
     {j_{\sigma(1)}\cdots j_{\sigma(k)}}
=
\operatorname{sgn}(\pi)\operatorname{sgn}(\sigma)
\rho^{i_1\cdots i_k}_{j_1\cdots j_k},
\label{eq:app-RDM-antisymmetry}
\end{equation}
where $\pi$ and $\sigma$ are permutations. Elements with repeated indices within either set, or with pair structures not covered above, vanish. With unrestricted sums over single-particle indices, the density matrices satisfy
\begin{equation}
\operatorname{Tr}\rho^{(k)}
=
\frac{N!}{(N-k)!},
\qquad k=1,\ldots,4.
\label{eq:app-RDM-traces}
\end{equation}

To characterize the irreducible densities, we evaluate their traces and Frobenius norms:
\begin{align}
\operatorname{Tr}\lambda^{(k)}
&=
\sum_{i_1,\ldots,i_k}
\lambda^{i_1\cdots i_k}_{i_1\cdots i_k},
\\
\left\|\lambda^{(k)}\right\|_{\mathrm F}
&=
\left[
\sum_{\substack{i_1,\ldots,i_k\\j_1,\ldots,j_k}}
\left|
\lambda^{i_1\cdots i_k}_{j_1\cdots j_k}
\right|^2
\right]^{1/2}.
\label{eq:lambda_trace_norm}
\end{align}
Each index runs over all $2L$ single-particle states, with no factorial normalization. Unlike the trace, which can be reduced by cancellations among diagonal elements, the Frobenius norm measures the overall tensor magnitude without such cancellations.

\begin{figure}
    \centering
    \includegraphics[width=0.8\linewidth]
    {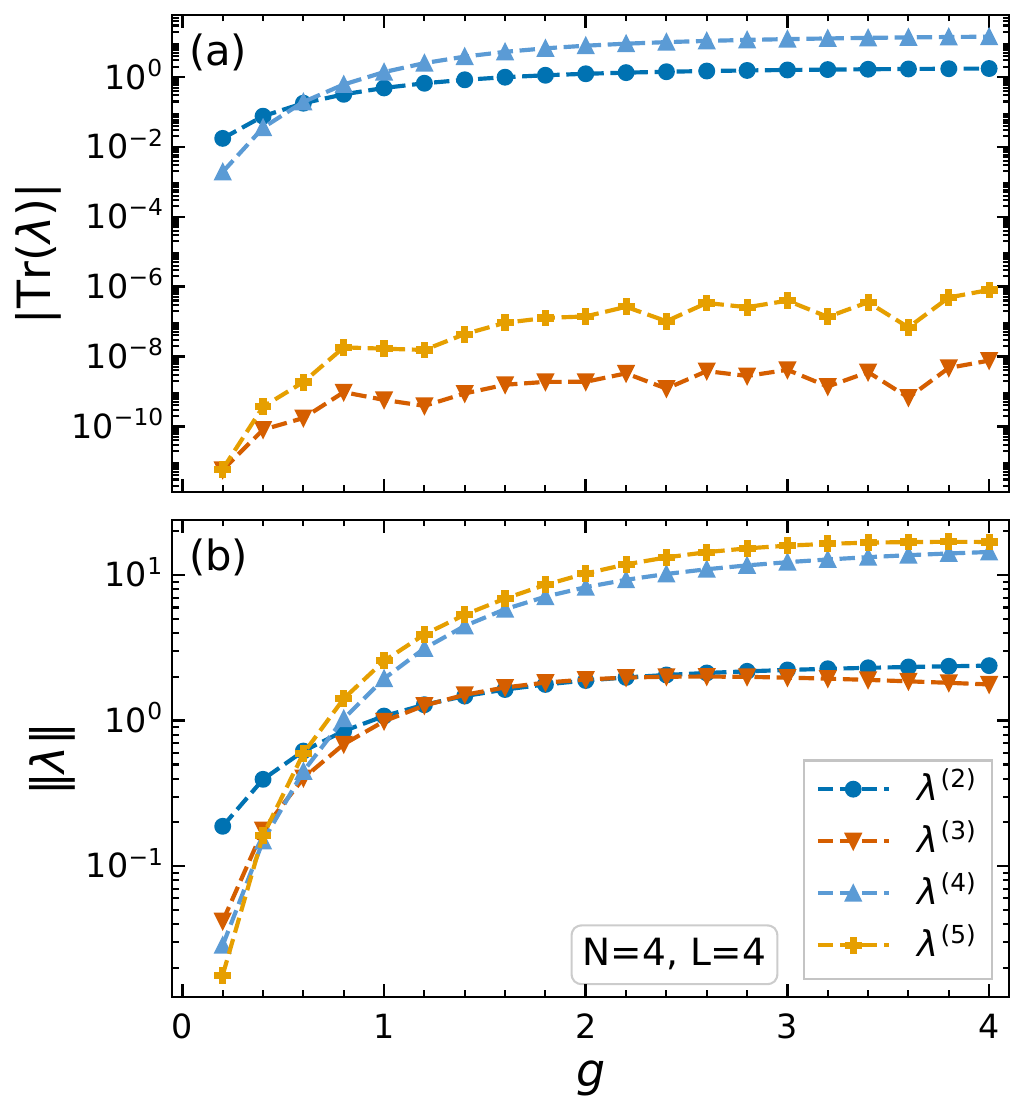}
    \caption{
    (a) Absolute traces and (b) Frobenius norms of the
    irreducible density matrices $\lambda^{(k)}$,
    $k=2,\ldots,5$, of the PNP-BCS reference as functions
    of the pairing strength $g$ in the P3H model with
    $N=4$ and $L=4$. At $g=0$, the reference reduces to
    the HF state, and all irreducible densities of rank
    $k\geq2$ vanish.}
    \label{fig:lambda_trace_norm}
\end{figure}

Figure~\ref{fig:lambda_trace_norm} shows these quantities for $k=2,\ldots,5$, with the absolute trace evaluated after summing the diagonal elements. The PNP-BCS variational space contains only paired configurations, within which the pair-breaking interaction has vanishing matrix elements. The optimized reference therefore depends on $g$ but not on $b$ in the Hamiltonian.

The traces in panel (a) exhibit pronounced even-odd staggering. At strong pairing, $|\operatorname{Tr}\lambda^{(2)}|$ is of order unity and $|\operatorname{Tr}\lambda^{(4)}|$ is of order $10$, whereas the odd-rank traces are approximately six to nine orders of magnitude smaller. These small traces do not imply uniformly small odd-rank tensor elements: panel (b) shows that all four Frobenius norms increase substantially with $g$. The norms of $\lambda^{(2)}$ and $\lambda^{(3)}$ reach values of order unity, while those of $\lambda^{(4)}$ and $\lambda^{(5)}$ reach values of order $10$. The small odd-rank traces thus reflect cancellations among diagonal elements. We also examine the traces of $\lambda^{(k)}$ for $N=2$ and $N=6$. It is shown that the magnitude of the trace of $\lambda^{(5)}$ exceeds that of $\lambda^{(4)}$ in the strong-pairing regime, $g>1.0$, where the norms of $\lambda^{(4,5)}$ are still overall larger than those of $\lambda^{(2,3)}$, see SM. Thus, the traces and norms of $\lambda^{(k)}$ depend on both the particle number and the pairing strength $g$.

The rapid growth of the norms of $\lambda^{(4)}$ and $\lambda^{(5)}$ in Fig.~\ref{fig:lambda_trace_norm}(b) before entering a slowly varying regime for $g\gtrsim2$, is consistent with the development and saturation of pairing correlations in this finite model space. Since the norms are not normalized by the number of tensor components, they do not directly establish the relative importance of different ranks in the flow equations. Their growth nevertheless suggests increased sensitivity to irreducible-density truncations at stronger pairing, where omitting higher-rank densities and their products can disrupt cancellations in the generator and flow equations. Retaining the complete set of required irreducible-density contributions is therefore important for both the accuracy of the final energy and the convergence of the MR-IMSRG flow.


%

\newpage 

\section*{Supplemental material}

\subsection*{Generators in the IMSRG}
\label{app:generators}

In the SR-IMSRG calculation, we employ the White generator~\cite{White2002}
\begin{align}
\eta^{\mathrm{W}}(s)
=&
\sum_{ph}
\frac{f^p_h(s)}{\Delta^p_h(s)}
\{A^p_h\}
+
\frac{1}{4}
\sum_{pp'hh'}
\frac{\Gamma^{pp'}_{hh'}(s)}
{\Delta^{pp'}_{hh'}(s)}
\{A^{pp'}_{hh'}\}
+\cdots-\mathrm{H.c.},
\label{eq:White-generator}
\end{align}
where particle orbitals are denoted by $pp'\ldots$, and hole orbitals by $hh'\ldots$.
The ellipsis denotes the analogous $3p3h$ and $4p4h$ contributions retained in SR-IMSRG(3) and SR-IMSRG(4), respectively. The subtraction of the Hermitian-conjugate terms ensures $\eta=-\eta^\dagger$.

To regularize the generator when an energy denominator becomes small, we also use the bounded arctangent variant~\cite{Hergert:2017PS}
\begin{align}
\eta^{\mathrm{atan}}(s)
=&
\frac{1}{2}\sum_{ph}
\arctan
\frac{2f^p_h(s)}{\Delta^p_h(s)}
\{A^p_h\}
+
\frac{1}{8}\sum_{pp'hh'}
\arctan
\frac{2\Gamma^{pp'}_{hh'}(s)}
{\Delta^{pp'}_{hh'}(s)}
\{A^{pp'}_{hh'}\} \nonumber \\
&+\cdots-\mathrm{H.c.}.
\label{eq:arctan-generator}
\end{align}

Both generators use M{\o}ller-Plesset (MP) denominators constructed from the diagonal matrix elements of the flowing one-body Hamiltonian~\cite{Hergert:2016PR,Heinz:2021},
\begin{equation}
\Delta^{p_1\cdots p_k}_{h_1\cdots h_k}(s) = \sum_{i=1}^{k}
f^{p_i}_{p_i}(s)
- \sum_{i=1}^{k}
f^{h_i}_{h_i}(s) = -\Delta_{p_1\cdots p_k}^{h_1\cdots h_k}(s).
\label{eq:MP-denominator-general}
\end{equation}
For example,
\begin{equation}
\Delta^p_h= f^p_p - f^h_h, \quad
\Delta^{pp'}_{hh'}
=
f^p_p+f^{p'}_{p'}
-f^h_h-f^{h'}_{h'}.
\label{eq:MP-denominators}
\end{equation}

We employ the Brillouin generator in the MR-IMSRG calculations. For MR-IMSRG(3), it is retained through the normal-ordered three-body (NO3B) level,
\begin{subequations}
\begin{align} 
\eta={}&\sum_{ij}\eta^i_j\{A^i_j\}
+\frac14\sum_{ijkl}\eta^{ij}_{kl}\{A^{ij}_{kl}\} +\frac1{36}\sum_{ijklmn}\eta^{ijk}_{lmn}\{A^{ijk}_{lmn}\},
\end{align}
\label{eq:app-H-eta}
\end{subequations}
with matrix elements defined by
\begin{equation}
\eta^{i_i \dots i_k}_{j_i \dots j_k}\equiv
\bra{\Phi_{\rm{ref}}}\left[H,\{A^{i_i \dots i_k}_{j_i \dots j_k} \}\right]\ket{\Phi_{\rm{ref}}}.
\end{equation}

We present below the one-, two-, and three-body matrix elements of the Brillouin generator, grouping their contributions by Hamiltonian component and irreducible density structure. In the notation $\eta^{(k)}(O,\lambda)$, with $O=f,\Gamma,W$ identifies the normal-ordered Hamiltonian component, while the $\lambda$ label specifies the irreducible density factors appearing explicitly in the expression. For example, $\eta^i_j(\Gamma,\lambda^{(2)})$ denotes the contribution involving the two-body Hamiltonian and the two-body irreducible density, whereas $\eta^{ij}_{kl}(W,\lambda^{(2)}\lambda^{(2)})$ involves the three-body Hamiltonian and a product of two two-body irreducible densities. Labels involving $\lambda^{(1)}$ denote contributions whose density dependence is expressed solely through the one-body occupations.

The one-body matrix element $\eta^i_j$ comprises three contributions,
\begin{equation}
\eta^i_j
=\eta^i_j(f,\lambda^{(1)})
+\eta^i_j(\Gamma,\lambda^{(2)})
+\eta^i_j(W,\lambda^{(3)}),
\label{eq:app-eta1-decomposition}
\end{equation}
where
\begin{subequations}
\begin{align}
\eta^i_j(f,\lambda^{(1)})
&=(n_j-n_i)f^j_i,
\label{eq:app-eta1-f}\\
\eta^i_j(\Gamma,\lambda^{(2)})
&=-\frac12\sum_{abc}
\left(
\Gamma^{ja}_{bc}\lambda^{ia}_{bc}
-\Gamma^{ab}_{ic}\lambda^{ab}_{jc}
\right),
\label{eq:app-eta1-Gamma}\\
\eta^i_j(W,\lambda^{(3)})
&=-\frac1{12}\sum_{abcde}
\left(
W^{dej}_{abc}\lambda^{ide}_{abc}
-W^{abc}_{dei}\lambda^{abc}_{jde}
\right).
\label{eq:app-eta1-W}
\end{align}
\end{subequations}

The two-body matrix element $\eta^{ij}_{kl}$ can be decomposed into eight terms,
\begin{align}
\eta^{ij}_{kl} 
&=\eta^{ij}_{kl}(f,\lambda^{(2)})
+\eta^{ij}_{kl}(\Gamma,\lambda^{(1)})
+\eta^{ij}_{kl}(\Gamma,\lambda^{(2)})
+\eta^{ij}_{kl}(\Gamma,\lambda^{(3)})\nonumber\\
&+\eta^{ij}_{kl}(W,\lambda^{(2)})
+\eta^{ij}_{kl}(W,\lambda^{(3)})
+\eta^{ij}_{kl}(W,\lambda^{(4)})
+\eta^{ij}_{kl}(W,\lambda^{(2)}\lambda^{(2)}).
\label{eq:app-eta2-decomposition}
\end{align}
where 
\begin{subequations}
\begin{align}
\eta^{ij}_{kl}(f,\lambda^{(2)})
=&\sum_a\left(
\hat{P}_{kl}f^l_a\lambda^{ij}_{ak}
-\hat{P}_{ij}f^a_j\lambda^{ai}_{kl}
\right),
\label{eq:app-eta2-f}
\end{align}

\begin{align}
\eta^{ij}_{kl}(\Gamma,\lambda^{(1)})
=&
(n_kn_l\bar n_i\bar n_j-\bar n_k\bar n_l n_i n_j)
\Gamma^{kl}_{ij},
\label{eq:app-eta2-Gamma-l1}
\end{align}

\begin{align}
\eta^{ij}_{kl}(\Gamma,\lambda^{(2)})  
&=\hat{P}_{ij}\hat{P}_{kl}\sum_{ab}(n_l-n_j)
\Gamma^{al}_{bj}\lambda^{ai}_{bk}\nonumber\\
&+\frac12\sum_{ab}
(n_kn_l-\bar n_k\bar n_l)
\Gamma^{kl}_{ab}\lambda^{ij}_{ab}\nonumber\\
&-\frac12\sum_{ab}
(n_in_j-\bar n_i\bar n_j)
\Gamma^{ab}_{ij}\lambda^{ab}_{kl}, 
\label{eq:app-eta2-Gamma-l2}
\end{align}

\begin{align}
\eta^{ij}_{kl}(\Gamma,\lambda^{(3)})
=\frac12\sum_{abc}
\left(
\hat{P}_{kl}\Gamma^{al}_{bc}\lambda^{aij}_{bck}
-\hat{P}_{ij}\Gamma^{bc}_{aj}\lambda^{bci}_{akl}
\right)
\end{align}

\begin{align}
\eta^{ij}_{kl}(W,\lambda^{(2)})
={}&\frac12\sum_{abc}\Big[
\hat{P}_{ij}(n_j\bar n_k\bar n_l+\bar n_jn_kn_l)
W^{ckl}_{abj}\lambda^{ic}_{ab}\nonumber\\
&
-\hat{P}_{kl}(n_l\bar n_i\bar n_j+\bar n_ln_in_j)
W^{abl}_{cij}\lambda^{ab}_{kc}
\Big],
\label{eq:app-eta2-W-l2}
\end{align}

\begin{align}
&\eta^{ij}_{kl}(W,\lambda^{(3)})\nonumber \\
=&\frac14\hat{P}_{ij}\hat{P}_{kl}
\sum_{abcd}(n_l-n_j)W^{abl}_{cdj}\lambda^{iab}_{kcd}\nonumber\\
+&\frac16\sum_{abcd}\Big[
(n_kn_l-\bar n_k\bar n_l)W^{dkl}_{abc}\lambda^{ijd}_{abc}
-(n_in_j-\bar n_i\bar n_j)W^{abc}_{dij}\lambda^{abc}_{kld}
\Big],
\label{eq:app-eta2-W-l3}
\end{align}

\begin{align}
\eta^{ij}_{kl}(W,\lambda^{(4)})
=&\frac1{12}\sum_{abcde}\left(
\hat{P}_{ij}W^{cde}_{abj}\lambda^{icde}_{klab}
-\hat{P}_{kl}W^{abl}_{cde}\lambda^{ijab}_{kcde}
\right),
\end{align}

\begin{align}
&\eta^{ij}_{kl}(W,\lambda^{(2)}\lambda^{(2)}) \nonumber\\
=&\frac12\sum_{abcde}\Big[
\frac12\hat{P}_{ij}W^{cde}_{abj}
\lambda^{cd}_{kl}\lambda^{ie}_{ab}
-\frac12\hat{P}_{kl}W^{abl}_{cde}
\lambda^{ij}_{cd}\lambda^{ab}_{ke}\nonumber\\
&
+\hat{P}_{ij}\hat{P}_{kl}W^{del}_{abc}
\lambda^{id}_{ka}\lambda^{je}_{bc}
-\hat{P}_{ij}\hat{P}_{kl}W^{abc}_{dej}
\lambda^{ia}_{kd}\lambda^{bc}_{le}
\Big].
\label{eq:app-eta2-W-l2l2}
\end{align}
\end{subequations}

The three-body matrix element $\eta^{ijk}_{lmn}$ contains many more terms: 
\begin{align}
\eta^{ijk}_{lmn}
={}&\eta^{ijk}_{lmn}(f,\lambda^{(3)})\nonumber\\
&+\eta^{ijk}_{lmn}(\Gamma,\lambda^{(2)})
+\eta^{ijk}_{lmn}(\Gamma,\lambda^{(3)})
+\eta^{ijk}_{lmn}(\Gamma,\lambda^{(2)}\lambda^{(2)})
+\eta^{ijk}_{lmn}(\Gamma,\lambda^{(4)})\nonumber\\
&+\eta^{ijk}_{lmn}(W,\lambda^{(1)})
+\eta^{ijk}_{lmn}(W,\lambda^{(2)})
+\eta^{ijk}_{lmn}(W,\lambda^{(3)})
+\eta^{ijk}_{lmn}(W,\lambda^{(4)})\nonumber\\
&+\eta^{ijk}_{lmn}(W,\lambda^{(2)}\lambda^{(2)})
+\eta^{ijk}_{lmn}(W,\lambda^{(5)})
+\eta^{ijk}_{lmn}(W,\lambda^{(2)}\lambda^{(3)}).
\label{eq:app-eta3-decomposition}
\end{align}
where the terms depending on $f$ and $\Gamma$ are given by
\begin{subequations}
\begin{align}
\eta^{ijk}_{lmn}(f,\lambda^{(3)})
={}&\sum_a\left[
\hat{P}(ij/k)f^a_k\lambda^{aij}_{lmn}
-\hat{P}(lm/n)f^n_a\lambda^{ijk}_{alm}
\right],  \\
\eta^{ijk}_{lmn}(\Gamma,\lambda^{(2)})
={}&\sum_a\hat{P}(ij/k)\hat{P}(mn/l)
(n_k\bar n_m\bar n_n+\bar n_kn_mn_n)
\Gamma^{mn}_{ak}\lambda^{ij}_{al}\nonumber\\
&-\sum_a\hat{P}(jk/i)\hat{P}(lm/n)
(n_n\bar n_j\bar n_k+\bar n_nn_jn_k)
\Gamma^{an}_{jk}\lambda^{ai}_{lm},\\
\eta^{ijk}_{lmn}(\Gamma,\lambda^{(3)})
={}&\sum_{ab}\hat{P}(ij/k)\hat{P}(lm/n)
(n_n-n_k)\Gamma^{an}_{bk}\lambda^{aij}_{blm}\nonumber\\
&+\frac12\sum_{ab}\Big[
\hat{P}(mn/l)(n_mn_n-\bar n_m\bar n_n)
\Gamma^{mn}_{ab}\lambda^{ijk}_{abl}\nonumber\\
& 
-\hat{P}(jk/i)(n_jn_k-\bar n_j\bar n_k)
\Gamma^{ab}_{jk}\lambda^{abi}_{lmn}
\Big],\\
\eta^{ijk}_{lmn}(\Gamma,\lambda^{(2)}\lambda^{(2)})
=&\frac12\sum_{abc}\Big[
\hat{P}(ij/k)\hat{P}(lm/n)\Gamma^{ab}_{ck}
\lambda^{ab}_{lm}\lambda^{ij}_{cn} \nonumber \\
&-\hat{P}(ij/k)\hat{P}(lm/n)\Gamma^{cn}_{ab}
\lambda^{ij}_{ab}\lambda^{ck}_{lm}\nonumber\\
& +2\hat{P}(jk/i)\hat{P}(l/m/n)\Gamma^{an}_{bc}
\lambda^{ai}_{bl}\lambda^{jk}_{cm} \nonumber \\
&-2\hat{P}(mn/l)\hat{P}(i/j/k)\Gamma^{bc}_{ak}
\lambda^{bi}_{al}\lambda^{cj}_{mn}
\Big],\\
\eta^{ijk}_{lmn}(\Gamma,\lambda^{(4)})
={}&\frac12\sum_{abc}\left[
\hat{P}(ij/k)\Gamma^{ab}_{ck}\lambda^{abij}_{clmn}
-\hat{P}(lm/n)\Gamma^{cn}_{ab}\lambda^{cijk}_{ablm}
\right]. 
\end{align}
\end{subequations}

The terms depending on $W$ are also decomposed in terms of different ranks of $\lambda^{(k)}$,
\beq 
\eta^{ijk}_{lmn}(W,\lambda^{(1)})
=(n_ln_mn_n\bar n_i\bar n_j\bar n_k
-\bar n_l\bar n_m\bar n_n n_in_jn_k)W^{lmn}_{ijk},
\eeq 
\beqn 
&&\eta^{ijk}_{lmn}(W,\lambda^{(2)})\nonumber\\
&&=\sum_{ab}\hat{P}(mn/l)\hat{P}(jk/i)
(n_mn_n\bar n_j\bar n_k-\bar n_m\bar n_n n_jn_k)
W^{amn}_{bjk}\lambda^{ai}_{bl}\nonumber\\
&&+\frac12\sum_{ab}\hat{P}(lm/n)
(n_n\bar n_i\bar n_j\bar n_k-\bar n_n n_in_jn_k)
W^{abn}_{ijk}\lambda^{ab}_{lm}\nonumber\\
&&-\frac12\sum_{ab}\hat{P}(ij/k)
(n_k\bar n_l\bar n_m\bar n_n-\bar n_k n_ln_mn_n)
W^{lmn}_{abk}\lambda^{ij}_{ab},
\label{eq:app-eta3-W-l2}
\eeqn
\beqn 
&&\eta^{ijk}_{lmn}(W,\lambda^{(3)})\nonumber\\
&=&\frac12\sum_{abc}\hat{P}(ij/k)\hat{P}(mn/l)
(n_k\bar n_m\bar n_n+\bar n_kn_mn_n)
W^{amn}_{bck}\lambda^{aij}_{bcl}\nonumber\\
&&-\frac12\sum_{abc}\hat{P}(jk/i)\hat{P}(lm/n)
(n_n\bar n_j\bar n_k+\bar n_nn_jn_k)
W^{bcn}_{ajk}\lambda^{bci}_{alm}\nonumber\\
&&+\frac16\sum_{abc}(n_in_jn_k+\bar n_i\bar n_j\bar n_k)
W^{abc}_{ijk}\lambda^{abc}_{lmn}\nonumber\\
&&-\frac16\sum_{abc}(n_ln_mn_n+\bar n_l\bar n_m\bar n_n)
W^{lmn}_{abc}\lambda^{ijk}_{abc},
\label{eq:app-eta3-W-l3}
\eeqn 
\beqn 
&&\eta^{ijk}_{lmn}(W,\lambda^{(4)})\nonumber\\
&=&\sum_{abcd}\Big[
\frac14\hat{P}(ij/k)\hat{P}(lm/n)(n_n-n_k)
W^{abn}_{cdk}\lambda^{abij}_{cdlm}\nonumber\\
&&+\frac16\hat{P}(mn/l)(n_mn_n-\bar n_m\bar n_n)
W^{amn}_{bcd}\lambda^{aijk}_{bcdl}\nonumber\\
&&-\frac16\hat{P}(jk/i)(n_jn_k-\bar n_j\bar n_k)
W^{bcd}_{ajk}\lambda^{bcdi}_{almn}
\Big],
\label{eq:app-eta3-W-l4} 
\eeqn 
\beqn 
&&\eta^{ijk}_{lmn}(W,\lambda^{(5)})\nonumber\\
&=&\frac1{12}\sum_{abcde}\left[
\hat{P}(ij/k)W^{abc}_{dek}\lambda^{abcij}_{delmn}
-\hat{P}(lm/n)W^{den}_{abc}\lambda^{deijk}_{abclm}
\right].\nonumber\\
\label{eq:app-eta3-W-l5}
\eeqn 
\begin{align}
&\eta^{ijk}_{lmn}(W,\lambda^{(2)}\lambda^{(2)})\nonumber\\
={}&\sum_{abcd}\Big[
\frac14\hat{P}(ij/k)\hat{P}(lm/n)(n_n-n_k)
W^{abn}_{cdk}\lambda^{ab}_{lm}\lambda^{ij}_{cd}\nonumber\\
&-\frac12\hat{P}(ij/k)\hat{P}(l/m/n)(n_n-n_k)
W^{abn}_{cdk}\lambda^{ab}_{cl}\lambda^{ij}_{dm}\nonumber\\
&-\frac12\hat{P}(lm/n)\hat{P}(i/j/k)(n_n-n_k)
W^{abn}_{cdk}\lambda^{ai}_{cd}\lambda^{bj}_{lm}\nonumber\\
&+\frac12\hat{P}(l/m/n)\hat{P}(i/j/k)(n_n-n_k)
W^{abn}_{cdk}\lambda^{ai}_{cl}\lambda^{bj}_{dm}
\Big]\nonumber\\
&+\frac12\hat{P}(jk/i)\hat{P}(mn/l)\sum_{abcd}\Big[
(n_mn_n-\bar n_m\bar n_n)W^{dmn}_{abc}
\lambda^{di}_{ab}\lambda^{jk}_{cl}\nonumber\\
&-(n_jn_k-\bar n_j\bar n_k)W^{abc}_{djk}
\lambda^{ab}_{dl}\lambda^{ci}_{mn}\nonumber\\
& 
+(n_mn_n-\bar n_m\bar n_n)W^{dmn}_{abc}
\lambda^{di}_{al}\lambda^{jk}_{bc}
-(n_jn_k-\bar n_j\bar n_k)W^{abc}_{djk}
\lambda^{ab}_{mn}\lambda^{ci}_{dl}
\Big],
\label{eq:app-eta3-W-l2l2}
\end{align}
and
\begin{align}
&\eta^{ijk}_{lmn}(W,\lambda^{(2)}\lambda^{(3)})\nonumber\\
={}&\sum_{abcde}\Bigg[
\frac12\hat{P}(jk/i)\hat{P}(l/m/n)
\left(W^{abn}_{cde}\lambda^{ai}_{cl}\lambda^{bjk}_{dem}\right)\nonumber\\
&-\frac12\hat{P}(i/j/k)\hat{P}(mn/l)
\left(W^{abc}_{dek}\lambda^{ai}_{dl}\lambda^{bcj}_{emn}\right)\nonumber\\
&+\frac14\hat{P}(i/j/k)\hat{P}(lm/n)
\left(W^{abc}_{dek}\lambda^{ai}_{lm}\lambda^{bcj}_{den}\right)\nonumber\\
&
+\frac12\hat{P}(jk/i)\hat{P}(lm/n)
\left(W^{abn}_{cde}\lambda^{ai}_{cd}\lambda^{bjk}_{elm}\right)\nonumber\\
&-\frac12\hat{P}(ij/k)\hat{P}(mn/l)
\left(W^{abc}_{dek}\lambda^{ab}_{dl}\lambda^{cij}_{emn}\right)\nonumber\\
&
-\frac14\hat{P}(ij/k)\hat{P}(l/m/n)
\left(W^{abn}_{cde}\lambda^{ij}_{cl}\lambda^{abk}_{dem}\right)\nonumber\\
&+\frac14\hat{P}(ij/k)\hat{P}(lm/n)
\left(W^{abc}_{dek}\lambda^{ab}_{lm}\lambda^{cij}_{den}\right)\nonumber\\
&
-\frac14\hat{P}(ij/k)\hat{P}(lm/n)
\left(W^{abn}_{cde}\lambda^{ij}_{cd}\lambda^{abk}_{elm}\right)\nonumber\\
&+\frac14\hat{P}(i/j/k)
\left(W^{abc}_{dek}\lambda^{ai}_{de}\lambda^{bcj}_{lmn}\right)\nonumber\\
&
+\frac16\hat{P}(jk/i)\hat{P}(lm/n)
\left(W^{abn}_{cde}\lambda^{ai}_{lm}\lambda^{bjk}_{cde}\right)\nonumber\\
&-\frac14\hat{P}(l/m/n)
\left(W^{abn}_{cde}\lambda^{ab}_{cl}\lambda^{ijk}_{dem}\right)\nonumber\\
&
-\frac16\hat{P}(ij/k)\hat{P}(mn/l)
\left(W^{abc}_{dek}\lambda^{ij}_{dl}\lambda^{abc}_{emn}\right)\nonumber\\
&+\frac1{12}\hat{P}(ij/k)
\left(W^{abc}_{dek}\lambda^{ij}_{de}\lambda^{abc}_{lmn}\right)\nonumber\\
&-\frac1{12}\hat{P}(lm/n)
\left(W^{abn}_{cde}\lambda^{ab}_{lm}\lambda^{ijk}_{cde}\right)
\Bigg].
\label{eq:app-eta3-W-l2l3}
\end{align}
 
Throughout these expressions, $P_{ij}$ exchanges $i$ and $j$ in all factors to its right. The permutation operators are defined as
\begin{subequations}
\begin{align}
\hat P_{ij}
&\equiv 1-P_{ij},\\
\hat P(ij/k)=\hat P(k/ij)
&\equiv 1-P_{ik}-P_{jk},\\
\hat P(i/j/k)
&\equiv 1-P_{ij}-P_{ik}-P_{jk}
\nonumber\\
&\quad+P_{ij}P_{ik}+P_{ik}P_{ij}.
\end{align}
\label{eq:app-permutation-definitions}
\end{subequations}
Here, $\hat P(ij/k)$(equivalently $\hat{P}(k/ij)$) completes the antisymmetrization of $k$ with an already antisymmetric pair $(i,j)$, while $\hat P(i/j/k)$ fully antisymmetrizes all three indices without assuming prior antisymmetry. 
Operators acting on lower indices are defined analogously.

Setting $\lambda^{(k)}=0$ for $k\geq2$ in the above expressions recovers the Brillouin generator for the SR-IMSRG(3)~\cite{Heinz:2021,Hergert:2017PS}. On the other hand, dropping all NO3B terms by setting $W=0$ and $\eta^{(3)}=0$ recovers the formulas for the MR-IMSRG(2) case~\cite{Hergert:2017PS,Yao:2018PRC}.

\subsection*{Additional numerical results}

\begin{figure}[bt]
    \centering
    \includegraphics[width=\linewidth]{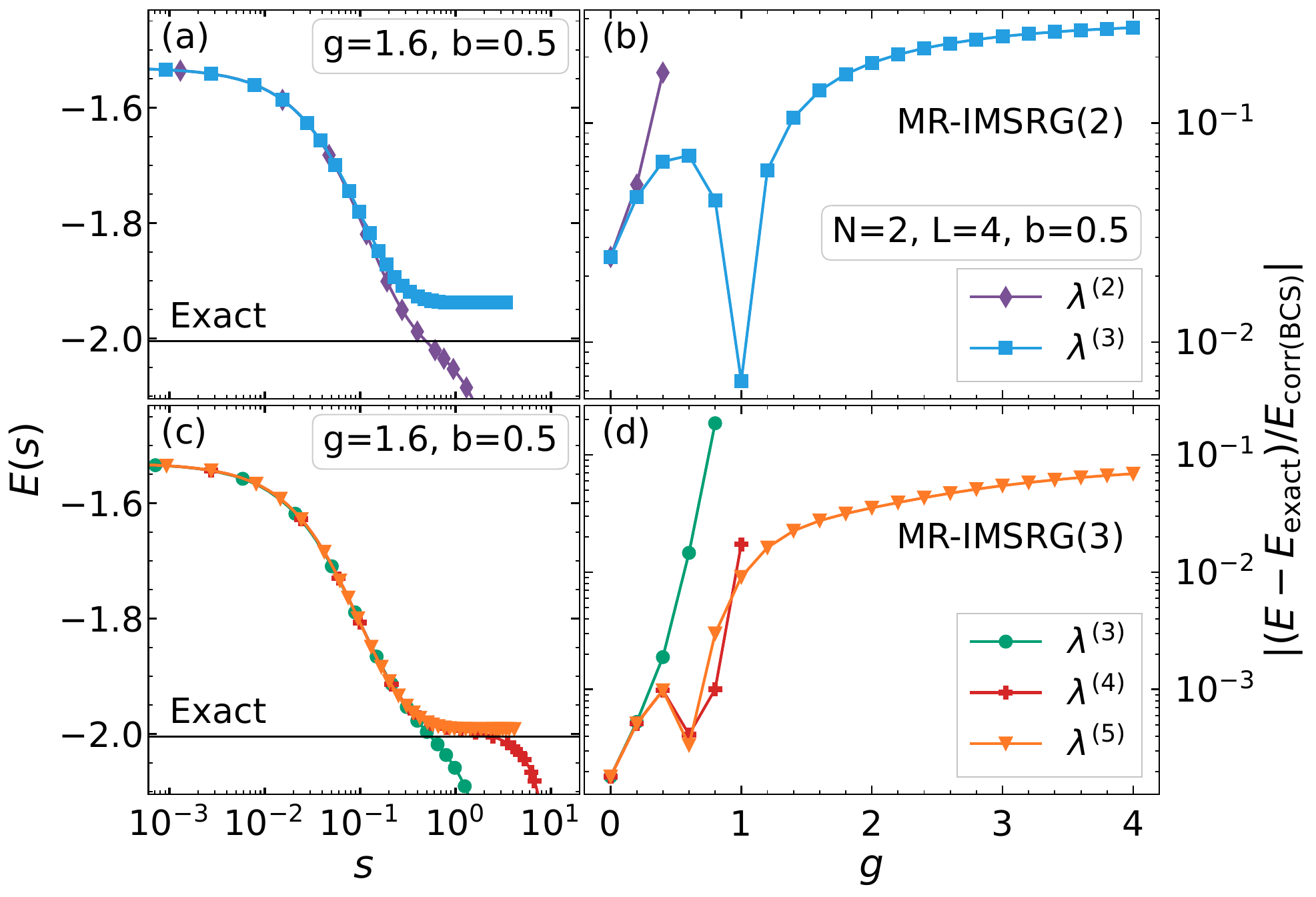}
    \caption{Ground-state energies as functions of the flow parameter $s$ [(a) and (c)] and normalized absolute energy errors as functions of the pairing strength $g$ [(b) and (d)] for the P3H model with $N=2$, $L=4$, and $b=0.5$. The upper and lower panels show MR-IMSRG(2) and MR-IMSRG(3) results, respectively, with irreducible density matrices retained through different ranks. The horizontal black lines in panels (a) and (c) indicate the exact ground-state energies. Missing data points indicate calculations that failed to converge. The dip in the normalized absolute error corresponds to a sign change in the energy error.}
    \label{fig:convergence_MRIMSRG_lambda_N2}
\end{figure}

\begin{figure}[bt]
    \centering
    \includegraphics[width=\linewidth]{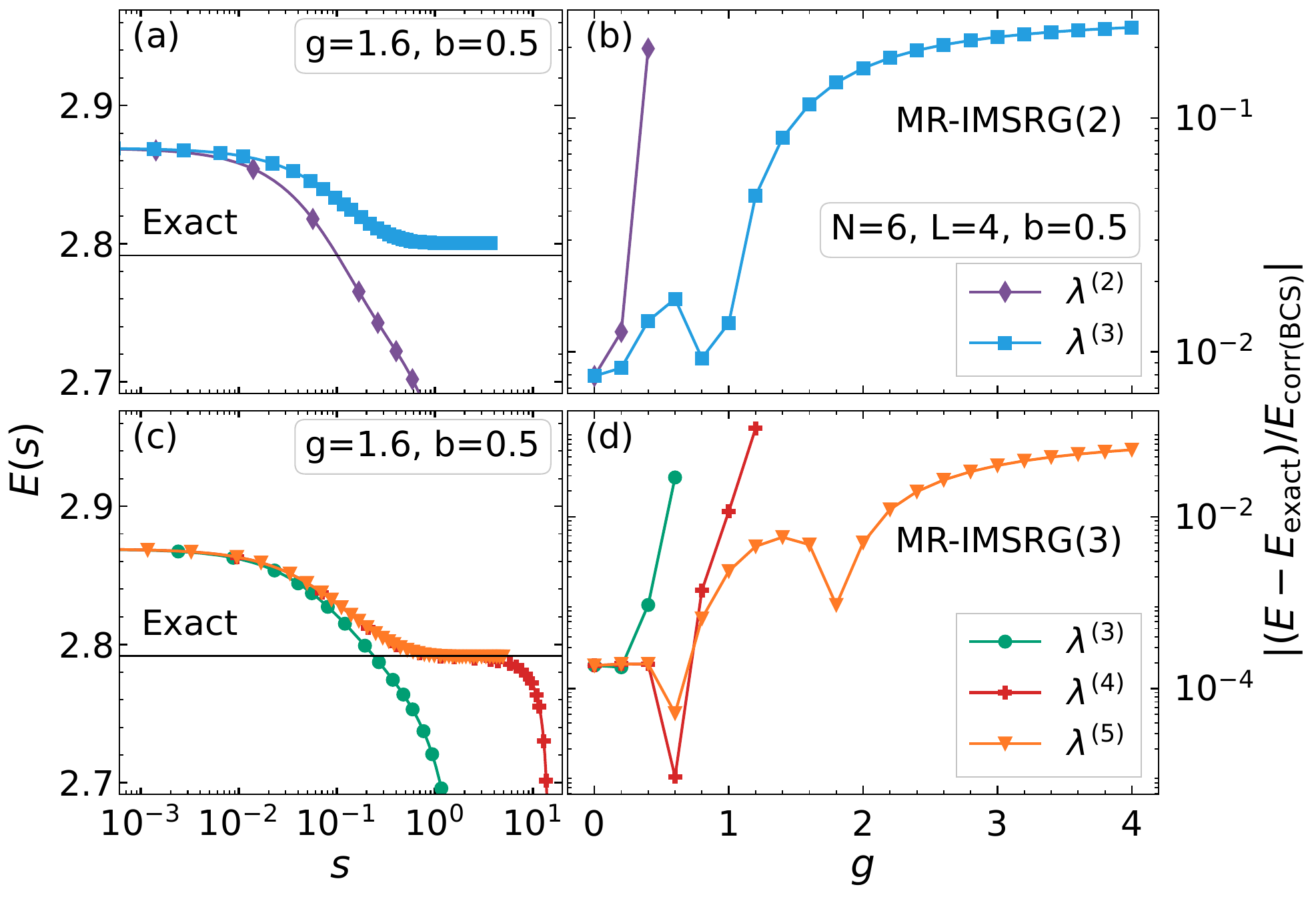}
    \caption{Same as Fig.~\ref{fig:convergence_MRIMSRG_lambda_N2}, but for $N=6$.}
    \label{fig:convergence_MRIMSRG_lambda_N6}
\end{figure}

\begin{figure}[bt]
    \centering
    \includegraphics[width=0.8\linewidth]{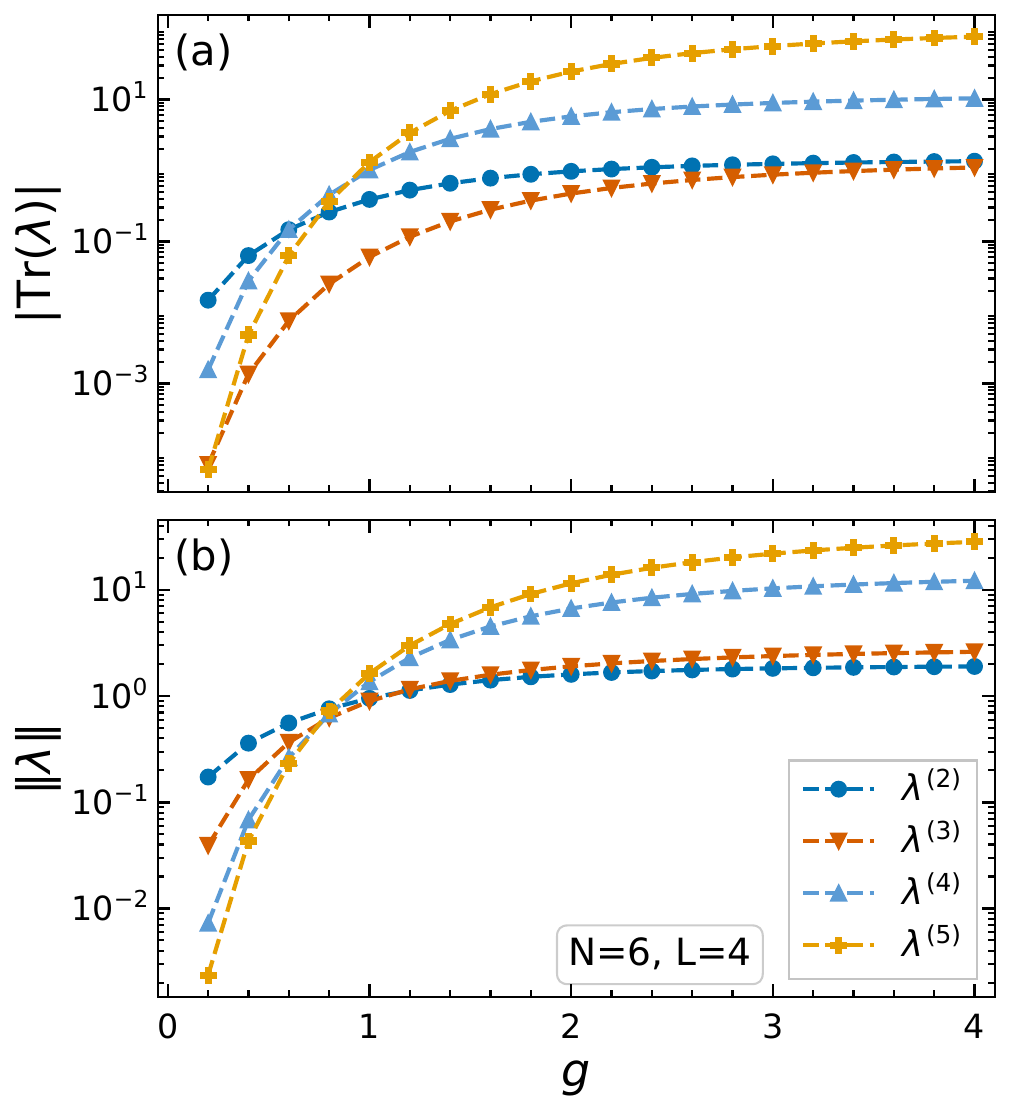}
    \caption{
    (a) Absolute traces and (b) Frobenius norms of the
    irreducible density matrices $\lambda^{(k)}$,
    $k=2,\ldots,5$, of the PNP-BCS reference state as functions
    of the pairing strength $g$ in the P3H model with
    $N=6$ and $L=4$. }
    \label{fig:lambda_trace_norm_N6}
\end{figure}

Figures~\ref{fig:convergence_MRIMSRG_lambda_N2} and~\ref{fig:convergence_MRIMSRG_lambda_N6} show the ground-state energies as functions of the flow parameter $s$ and the normalized energy errors as functions of the pairing strength $g$ for $N=2$ and $N=6$, respectively. The MR-IMSRG(2) and MR-IMSRG(3) results are obtained with irreducible density matrices retained through different ranks. For both particle numbers, the results exhibit the same qualitative trends as those for $N=4$ discussed in the main text, highlighting the importance of higher-rank irreducible densities for both flow convergence and energy accuracy. Their importance becomes more pronounced with increasing pairing strength, consistent with the growth in the magnitudes of their trace and their Frobenius norms shown in Fig.~\ref{fig:lambda_trace_norm_N6}.

Figure~\ref{fig:lambda_trace_norm_N6} shows (a) the absolute traces and (b) the Frobenius norms of the irreducible density matrices $\lambda^{(k)}$ for $N=6$ and $L=4$. We have also verified that these quantities coincide with those for $N=2$ and $L=4$, consistent with the particle-hole correspondence between the two systems. In the strong-pairing regime, $g>1.0$, the absolute trace of $\lambda^{(5)}$ exceeds that of $\lambda^{(4)}$ by approximately one order of magnitude. The Frobenius norms of $\lambda^{(4)}$ and $\lambda^{(5)}$ are also generally about one order of magnitude larger than those of $\lambda^{(2)}$ and $\lambda^{(3)}$. These results, together with the results in the main text, illustrate the dependence of the traces and norms on both the particle number and the pairing strength $g$.


\begin{figure}[bt]
\centering
\includegraphics[width=0.8\columnwidth]{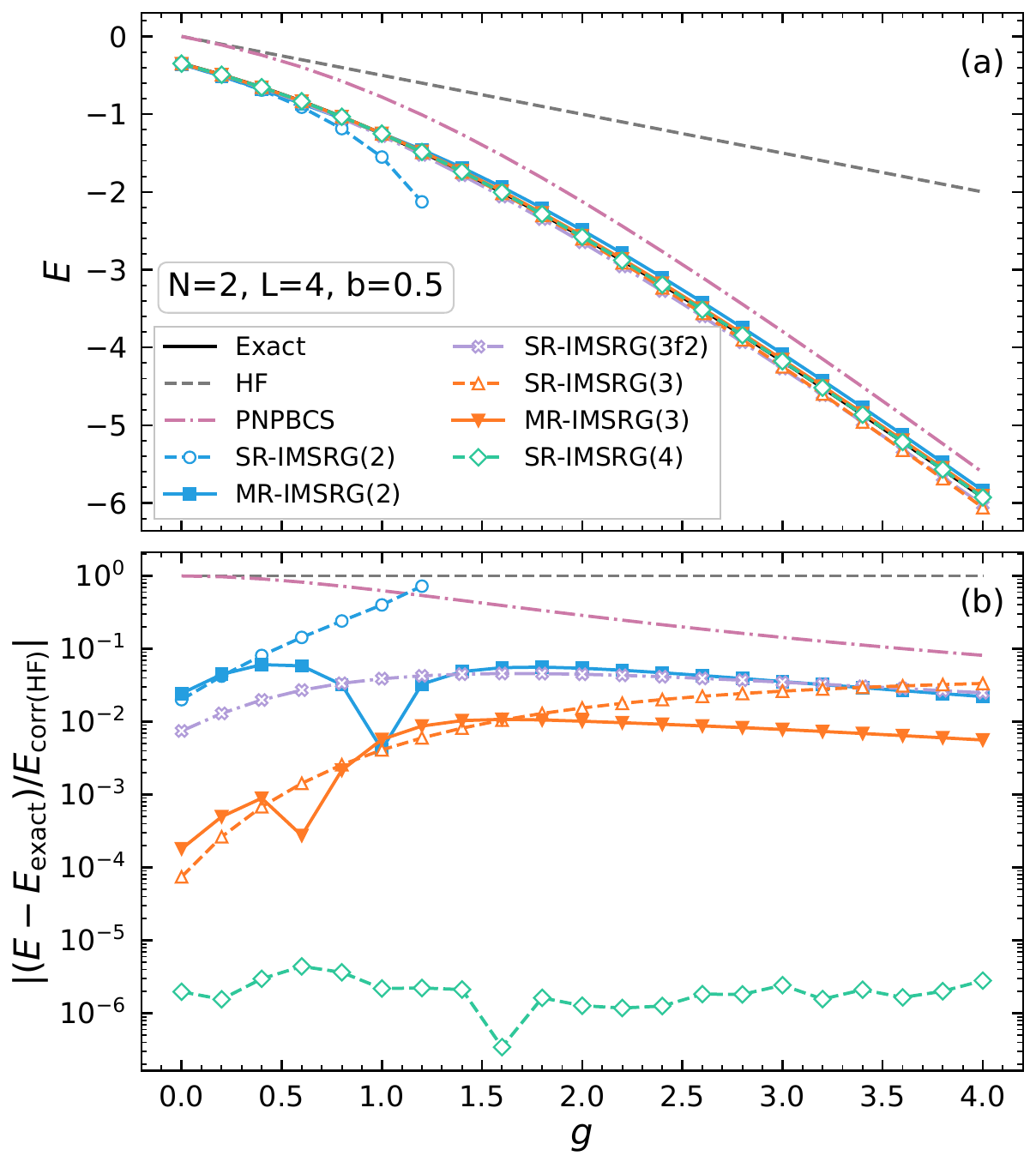}
\caption{(a) Ground-state energies obtained using HF, PNP-BCS(VAP), SR-IMSRG($n$), MR-IMSRG($n$), and FCI (exact) as functions of the pairing strength $g$ for the P3H model with $N=2$, $L=4$, and $b=0.5$. (b) Corresponding absolute energy errors normalized
by the HF correlation energy.
}
\label{fig:MRvsSR_N2_b0.5}
\end{figure}

\begin{figure}[bt]
\centering
\includegraphics[width=0.8\columnwidth]{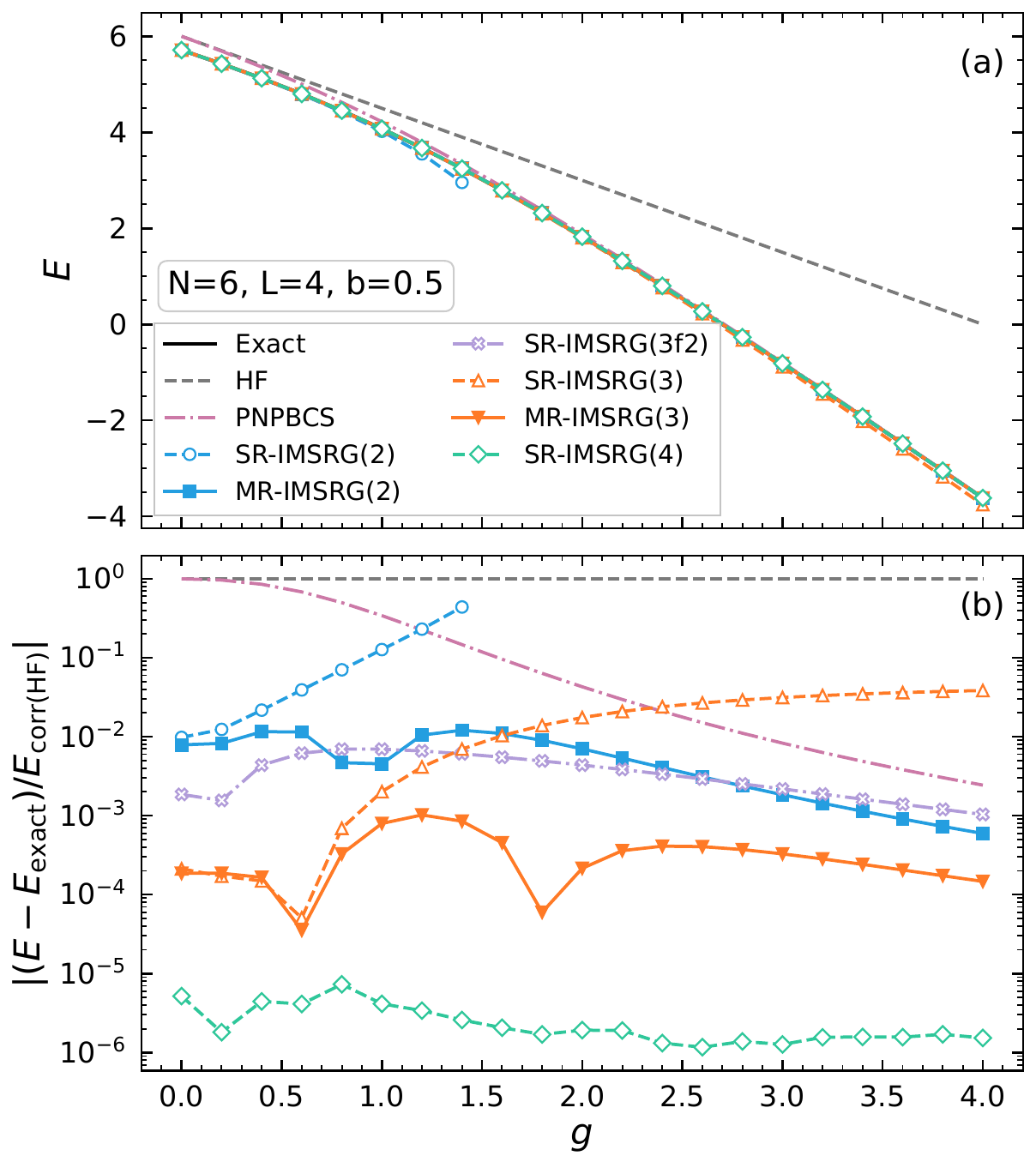}
\caption{Same as Fig.~\ref{fig:MRvsSR_N2_b0.5}, but for $N=6$.}
\label{fig:MRvsSR_N6_b0.5}
\end{figure}


\begin{figure}[bt]
    \centering
    \includegraphics[width=0.8\linewidth]{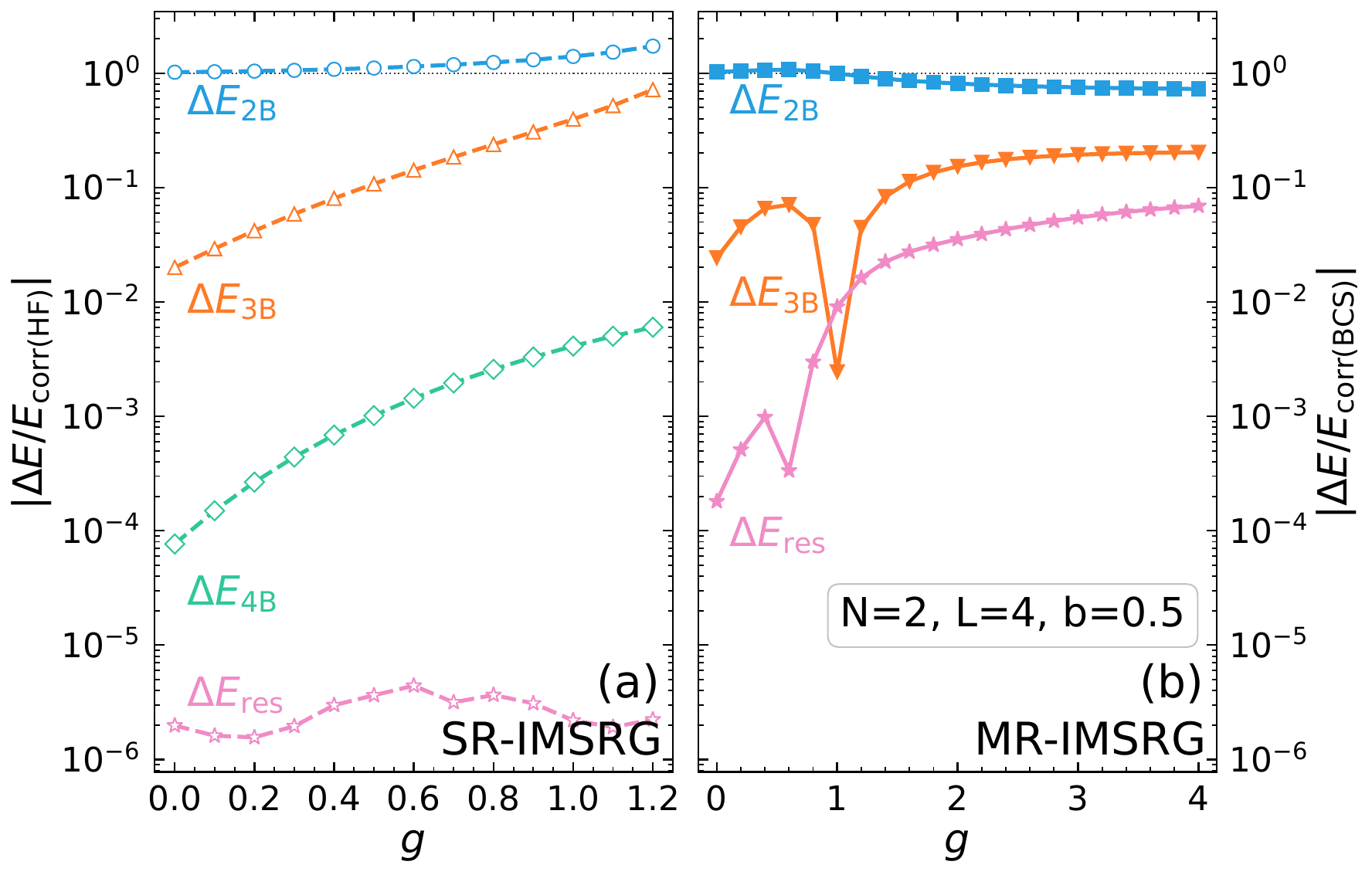}
    \caption{
    Successive energy corrections and residuals, as defined in the main text, for the P3H model with $N=2$, $L=4$, and $b=0.5$. (a) SR-IMSRG results through NO4B, normalized by the HF correlation energy. (b) MR-IMSRG results through NO3B, normalized by the PNP-BCS correlation energy.}
    \label{fig:SR_MR_truncation_N2_b0.5}
\end{figure}

\begin{figure}[bt]
    \centering
    \includegraphics[width=0.8\linewidth]{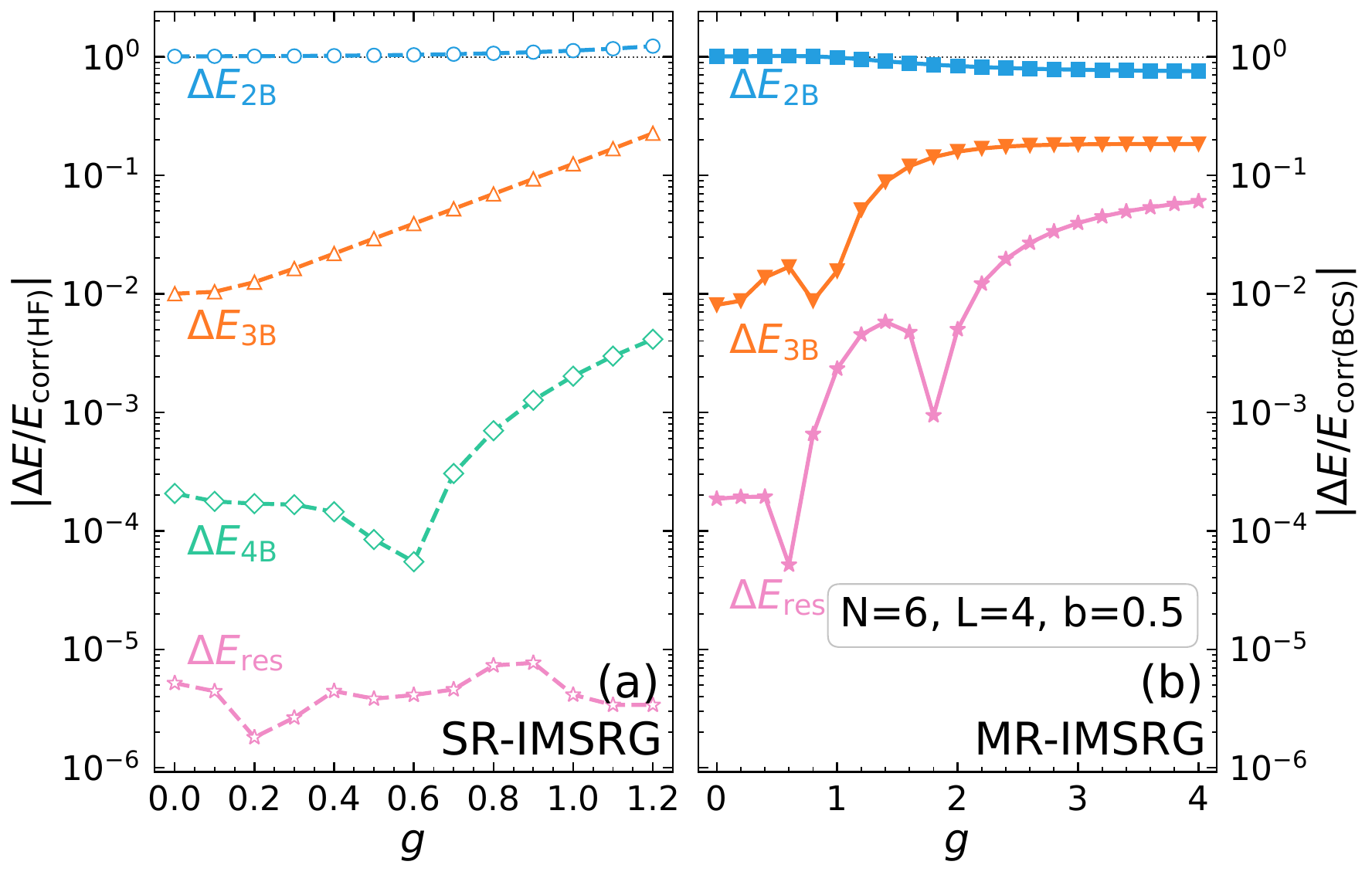}
    \caption{Same as Fig.~\ref{fig:SR_MR_truncation_N2_b0.5}, but for $N=6$.}
    \label{fig:SR_MR_truncation_N6_b0.5}
\end{figure}

Figures~\ref{fig:MRvsSR_N2_b0.5} and~\ref{fig:MRvsSR_N6_b0.5} compare the ground-state energies and normalized absolute energy errors obtained using HF, PNP-BCS, SR-IMSRG($n$), MR-IMSRG($n$), and FCI for $N=2$ and $N=6$, respectively. Figures~\ref{fig:SR_MR_truncation_N2_b0.5} and~\ref{fig:SR_MR_truncation_N6_b0.5} show the corresponding successive energy corrections and residuals defined in the main text, with SR-IMSRG and MR-IMSRG results presented in panels (a) and (b), respectively. These figures follow the same format and normalization conventions as the $N=4$ results in the main text, with $L=4$ and $b=0.5$ unchanged.

For both $N=2$ and $N=6$, the results are qualitatively consistent with those for $N=4$. At the same normal-ordered $n$-body truncation level, MR-IMSRG($n$) yields more accurate energies than SR-IMSRG($n$) for $n=2$ and $n=3$, particularly in the strong-pairing regime. Retaining higher-rank operators generally improves the accuracy, with successively smaller energy corrections within the convergence ranges examined.
A notable feature of the P3H model with $L=4$ is that SR-IMSRG(4) reproduces the exact energies for both $N=2$ (two particles) and $N=6$ (two holes). For an $A$-particle system with an $A$-particle single-determinant reference, the wave function contains at most $A$-particle--$A$-hole excitations. The normal-ordered $n$-body truncation is therefore exact for $n\geq 2A$~\cite{Stroberg:2024}. The same argument applies to an $A$-hole system in the hole representation. The residual error, of order $10^{-6}$, is attributed to the flow termination criterion.

These additional benchmarks support the conclusions drawn in the main text and indicate that the observed behavior extends beyond the $N=4$ system.


\begin{figure}[bt]
\centering
\includegraphics[width=0.8\columnwidth]{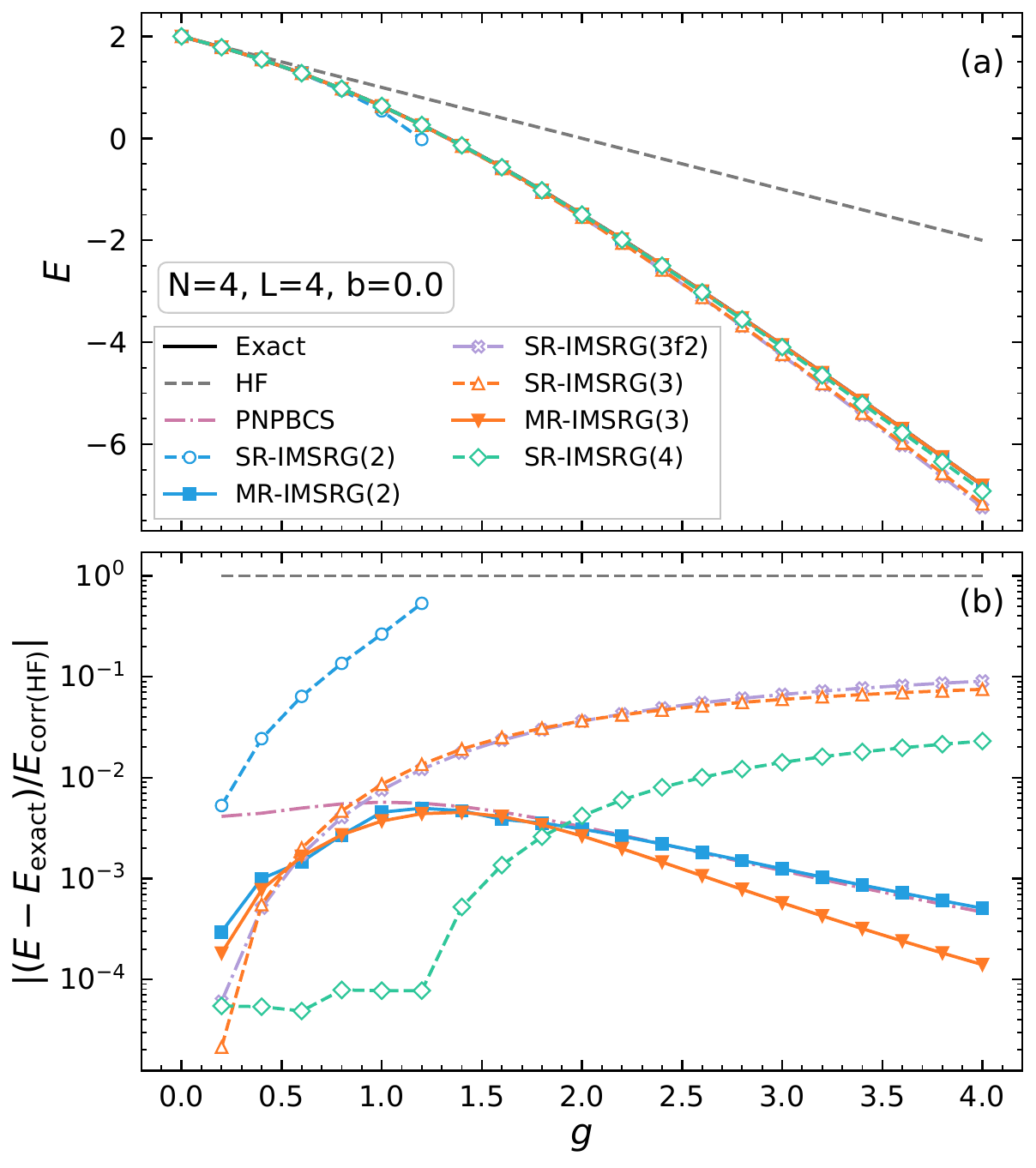}
\caption{(a) Ground-state energy as a function of the pairing strength $g$ from different calculations, including Hartree-Fock (HF), PNP-BCS (VAP), SR-IMSRG(n), MR-IMSRG(n), and FCI(Exact). (b) Corresponding absolute deviations from the exact energy, normalized by the HF correlation energy. 
}
\label{fig:MRvsSR_N4_b0.0}
\end{figure}

Figure~\ref{fig:MRvsSR_N4_b0.0} compares the ground-state energies of the $N=4$ P3H system with $b=0$ as functions of the pairing strength $g$, obtained using HF, PNP-BCS (VAP), SR-IMSRG($n$), MR-IMSRG($n$), and FCI (exact). It also shows the corresponding absolute energy errors normalized to the HF correlation energy. The results exhibit trends similar to those found for $b=0.5$, and the conclusions drawn in the main text remain valid.

\begin{figure}[bt]
\centering
\includegraphics[width=\columnwidth]{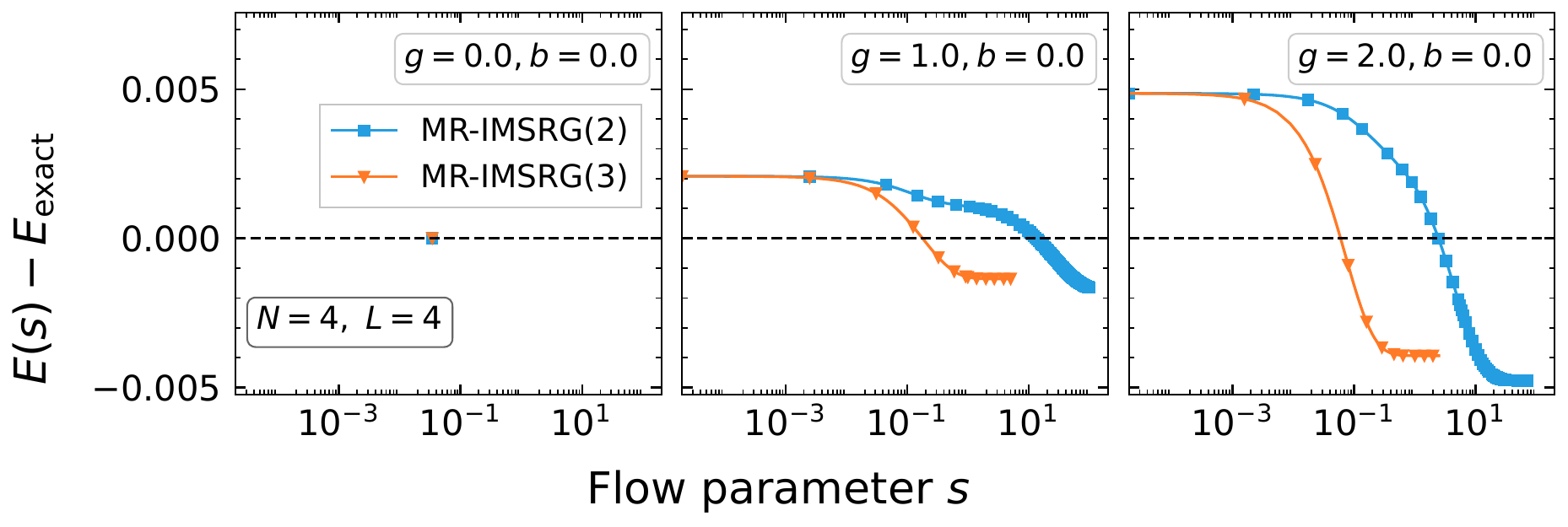}
\caption{The energy evolution with the flow parameter in the MR-IMSRG(2) and MR-IMSRG(3) for the cases with different interaction strengths $g$ at $b=0.0$ in the P3H model with $N=4$ and $L=4$.
}
\label{fig:MR-IMSRG2_3_flow_b0.0}
\end{figure}

\begin{figure}[bt]
\centering
\includegraphics[width=\columnwidth]{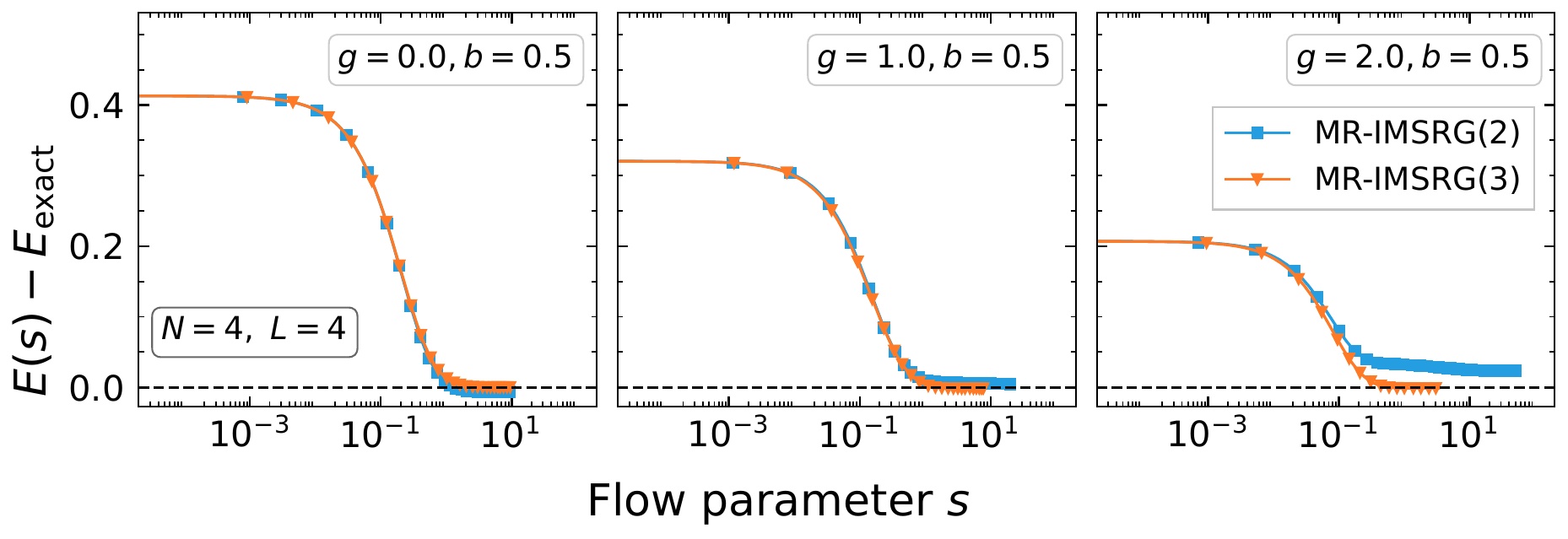}
\caption{Same as Fig.~\ref{fig:MR-IMSRG2_3_flow_b0.0},
but for $b=0.5$.
}
\label{fig:MR-IMSRG2_3_flow_b0.5}
\end{figure}

Figures~\ref{fig:MR-IMSRG2_3_flow_b0.0} and~\ref{fig:MR-IMSRG2_3_flow_b0.5} show the evolution of the ground-state energy with the flow parameter $s$ in MR-IMSRG(2) and MR-IMSRG(3) for the different pairing strengths $g$ at $b=0$ and $b=0.5$, respectively. For $g=b=0$, the reference state is exact, and no flow occurs. As the pairing strength $g$ increases, the converged energies deviate increasingly from the exact values. In all cases considered, MR-IMSRG(3) yields smaller errors than MR-IMSRG(2), demonstrating the improved accuracy obtained by retaining normal-ordered three-body operators.

\end{document}